\documentclass[lettersize,journal]{IEEEtran}
\usepackage{amsmath}
\usepackage{amsthm}
\usepackage{amsfonts}
\usepackage{amssymb}
\usepackage{mathtools}
\usepackage{algorithmic}
\usepackage{algorithm}
\usepackage{array}
\usepackage{pifont}
\usepackage[table]{xcolor}
\usepackage{tabularx}
\usepackage{makecell}
\usepackage{multirow}
\usepackage[compact]{titlesec}
\usepackage[caption=false,font=normalsize,labelfont=sf,textfont=sf]{subfig}
\usepackage{textcomp}
\usepackage{stfloats}
\usepackage{url}
\usepackage{verbatim}
\usepackage{graphicx}
\usepackage{cite}
\usepackage[hidelinks]{hyperref}
\hypersetup{
  colorlinks = true, 
  citecolor = black, 
  linkcolor = black, 
  urlcolor = black
}

\newcommand{\xmark}{\ding{53}}
\definecolor{gray}{HTML}{E0E0E0}

\newcommand{\rw}[1]{\textcolor{black}{#1}}

\newcommand{\RNum}[1]{\uppercase\expandafter{\romannumeral #1\relax}}
\DeclareMathOperator*{\argmax}{argmax}

\newtheorem{definition}{Definition}
\newtheorem{theorem}{Theorem}

\begin{document}

\title{Heterogeneous Domain Adaptation for IoT Intrusion Detection: A Geometric Graph Alignment Approach}

\author{Jiashu Wu,~\IEEEmembership{Graduate Student Member,~IEEE},
        Hao Dai,~\IEEEmembership{Graduate Student Member,~IEEE},
        Yang Wang,~\IEEEmembership{Member,~IEEE}\thanks{* Yang Wang is the corresponding author},
        Kejiang Ye,~\IEEEmembership{Member,~IEEE},
        and Chengzhong Xu,~\IEEEmembership{Fellow,~IEEE}% <-this % stops a space
\thanks{Jiashu Wu, Hao Dai, Yang Wang and Kejiang Ye are with Shenzhen Institute of Advanced Technology, Chinese Academy of Sciences, Shenzhen 518055, China. Email: \{js.wu, hao.dai, yang.wang1, kj.ye\}@siat.ac.cn}% <-this % stops a space
\thanks{Jiashu Wu and Hao Dai are also with University of Chinese Academy of Sciences, Beijing 100049, China. Email: \{wujiashu21, daihao19\}@mails.ucas.ac.cn}% <-this % stops a space
\thanks{Chengzhong Xu is with the State Key Laboratory of IoT for Smart City, Faculty of Science and Technology, University of Macau, Macau 999078, China. Email: czxu@um.edu.mo}% <-this % stops a space
\thanks{Manuscript received January 00, 2022; revised January 00, 2023.}
\thanks{Copyright (c) 2023 IEEE. Personal use of this material is permitted. However, permission to use this material for any other purposes must be obtained from the IEEE by sending a request to pubs-permissions@ieee.org. }}

\markboth{IEEE Internet of Things Journal}%
{Wu \MakeLowercase{\textit{et al.}}: Heterogeneous Domain Adaptation for IoT Intrusion Detection: A Geometric Graph Alignment Approach}


\maketitle

\begin{abstract}
Data scarcity hinders the usability of data-dependent algorithms when tackling IoT intrusion detection (IID). To address this, we utilise the data rich network intrusion detection (NID) domain to facilitate more accurate intrusion detection for IID domains. In this paper, a Geometric Graph Alignment (GGA) approach is leveraged to mask the geometric heterogeneities between domains for better intrusion knowledge transfer. Specifically, each intrusion domain is formulated as a graph where vertices and edges represent intrusion categories and category-wise interrelationships, respectively. The overall shape is preserved via a confused discriminator incapable to identify adjacency matrices between different intrusion domain graphs. A rotation avoidance mechanism and a centre point matching mechanism is used to avoid graph misalignment due to rotation and symmetry, respectively. Besides, category-wise semantic knowledge is transferred to act as vertex-level alignment. To exploit the target data, a pseudo-label election mechanism that jointly considers network prediction, geometric property and neighbourhood information is used to produce fine-grained pseudo-label assignment. Upon aligning the intrusion graphs geometrically from different granularities, the transferred intrusion knowledge can boost IID performance. Comprehensive experiments on several intrusion datasets demonstrate state-of-the-art performance of the GGA approach and validate the usefulness of GGA constituting components. 
\end{abstract}

\begin{IEEEkeywords}
Internet of Things (IoT), Intrusion Detection, Domain Adaptation, Geometric Graph Alignment, Pseudo Label Election
\end{IEEEkeywords}

\section{Introduction}\label{sec:introduction}

\IEEEPARstart{I}{n}ternet of Things (IoT) devices become indispensable for various real world applications and innovatively transform several fields such as healthcare \cite{bhuiyan2021internet,9933783}, etc. However, limited power and computational capability of IoT devices hinder the applicability of powerful security mechanisms, together with relatively infrequent maintenance, making IoT vulnerable to malicious intrusions. Therefore, to keep the IoT infrastructure safe, an effective IoT intrusion detection (IID) system is vital. To advance the intrusion detection techniques, several research directions become popular. Dietz \cite{dietz2018iot} proposed to automatically scan IoT devices for pre-defined vulnerability patterns, and isolated suspicious devices to block the botnet spreading. McDermott \cite{mcdermott2018botnet} tackled the problem via deep recurrent neural network (RNN) and achieved satisfying detection performance. However, these efforts required either a thorough intrusion pattern repository, or abundant labelled training data, which is expensive to collect and time-consuming to annotate, and is especially difficult for IoT due to factors such as data privacy concerns, the frequent emergence of new IoT things, etc. Therefore, the data-scarcity of IoT hinders the usability of these rule or data-dependent methods. 

\begin{figure}[t]
  \centering
  \includegraphics[width=0.48\textwidth]{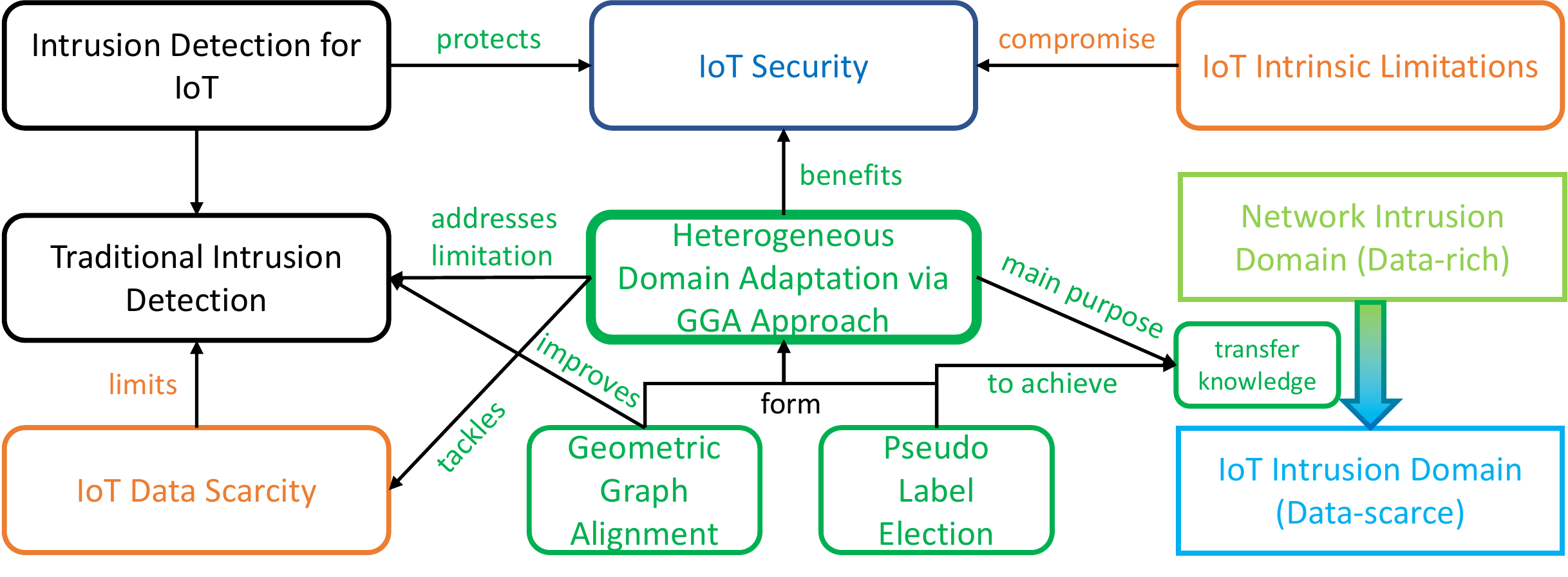}
  \caption{\rw{The general motivation of the GGA approach}}
  \label{fig:intuition}
  \vspace{-0.5cm}
\end{figure}

Considering that the Internet intrusion data is richer than IoT domains, and they share several similar intrusion categories, several domain adaptation-based (DA) methods were proposed to treat the network intrusion (NI) as the source domain, and transfer rich intrusion knowledge to facilitate the data-scarce target IoT intrusion (II) domains. By achieving domain-invariant alignment, the transferred intrusion knowledge can facilitate more accurate IID. For instance, Ning \cite{ning2021malware} presented a Laddernet-based DA solution to improve the intrusion classification accuracy and secure the industrial IoT infrastructures. Hu \cite{hu2022deep} studied a deep subdomain adaptation network with attention mechanism and focused on distribution alignment between domains via local maximum mean discrepancy. Methods such as \cite{wang2020prototype,3482115} proposed to achieve intrusion domain alignment by aligning the graph learning results. Efforts such as \cite{hsieh2016recognizing,li2021cross,yao2019heterogeneous} attempted to explore unlabelled target domain via directly predicted, threshold selected or softly assigned pseudo-labels (PLs), respectively, to facilitate better intrusion knowledge transfer. 

\begin{figure}[t]
    \centering
    \includegraphics[width=0.48\textwidth]{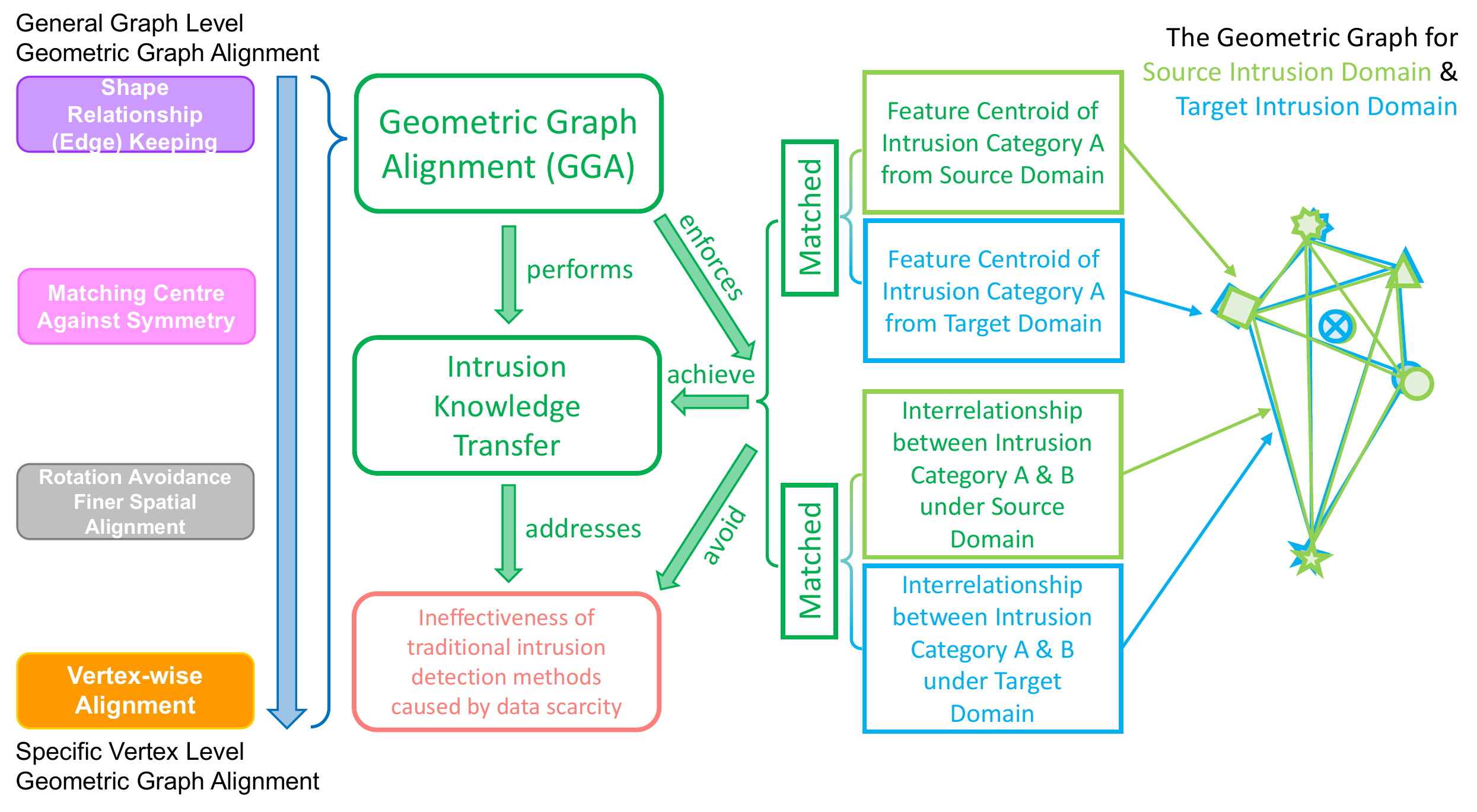}
    \caption{\rw{An overview of the geometric graph alignment (GGA) mechanism}}
    \label{fig:gga_motivation}
    \vspace{-0.5cm}
\end{figure}

Despite their success, they left some deficiencies that need to be addressed. Previous DA-based ID methods didn't tackle the problem from a geometric graph perspective and failed to explore unlabelled target domain via geometric and neighbourhood-aware PLs. The ignorance of intrinsic geometric properties in domain graphs and the under-explored target domain can result in coarse-grained alignment. Although some graph-based DA methods were proposed, they didn't attempt graph alignment from a pure geometric perspective, leaving the geometric properties under-explored. Despite some methods can partially convey the geometric properties through graph embedding, however, the embedding learning is highly data-dependent which is challenging for IoT scenarios. Besides, although there were some traditional PL-based DA methods, their isolated PL assignment strategy failed to leverage the geometric and neighbourhood information between labels, which may produce error-prone PLs and mislead the intrusion knowledge transfer. 

To address these limitations and achieve more fine-grained intrusion knowledge transfer, \rw{following the motivation illustrated in Fig. \ref{fig:intuition} and \ref{fig:gga_motivation}}, we propose a Geometric Graph Alignment (GGA) approach that works under the semi-supervised heterogeneous domain adaptation (HDA) setting, i.e., the target is scarcely-labelled and significant source-target heterogeneities exist, such as having diverse feature spaces, following different distributions, etc. To positively exploit the unlabelled target domain, we utilise a pseudo-label election (PLE) mechanism. To prevent the error-prone PL from misleading the model, the geometric property is considered to eliminate confident but incorrect PLs based on their geometric relationship with each category. Besides, the PLE consults the neighbourhood label information when assigning PLs to avoid near-boundary ambiguous PLs, which cannot be fulfilled by traditional isolated PL assignment strategies. By jointly considering the network prediction, the geometric property and the neighbourhood information, the PLE can boost pseudo label accuracy and hence lead to positive intrusion knowledge transfer. 

The GGA then formulates each domain as a graph, where vertices and edges represent intrusion categories and their interrelationships. As illustrated in Fig. \ref{fig:gga_motivation}, enforcing a perfect geometric graph alignment can have each intrusion category and their interrelationships well aligned between domains. Firstly, with the help of the PLE, the GGA performs a graph-level shape keeping via a confused discriminator which is incapable to distinguish weighted adjacency matrices between intrusion domain graphs. Upon aligning the graph shapes, a centre point matching mechanism and a rotation avoidance mechanism avoid graph misalignment caused by symmetry and graph rotation, respectively. Finally, the GGA will perform a vertex-level matching by preserving categorical correlation knowledge between domains, which equivalently aligns the graph vertex of each intrusion category between domains. Holistically, they form a graph alignment framework from general to specific level from a geometric perspective. The GGA can robustly transfer the enriched intrusion knowledge from the NI domain to facilitate more accurate intrusion detection in the II domain and hence secure IoT infrastructures. \rw{GGA's motivation has been illustrated in Fig. \ref{fig:intuition}-\ref{fig:gga_motivation}. }

In summary, the contributions of this paper are three-fold:
\begin{itemize}
    \item We propose to transfer enriched intrusion knowledge from the NI domain to facilitate more accurate intrusion detection for data-scarce IoT domains and formulate it as a semi-supervised HDA problem. 
    \item To our best knowledge, we are the pioneer to tackle this HDA problem from a pure geometric graph alignment perspective with the help of the PLE mechanism. \rw{Rather than using isolated coarse-grained PL strategies, }the PLE makes fine-grained PL assignment by jointly considering geometric and neighbourhood information \rw{to filter confident but geometrically incorrect PLs} and near-boundary ambiguous PLs. The GGA then aligns domain graphs geometrically through four mechanisms, holistically forming a graph alignment framework from graph to vertex granularity. 
    \item We conduct comprehensive experiments of several tasks on $5$ widely used intrusion detection datasets to verify the superior performance achieved by the GGA, and show the usefulness of its constituting components. 
\end{itemize}

The rest of the paper is organised as follows: Section \ref{sec:related_work} presents related works by categories and demonstrates the research opportunities of the GGA method. Section \ref{sec:model_preliminary_and_architecture} provides model preliminaries, graph formulations and \rw{the GGA model architecture. }The detailed geometric graph alignment and pseudo label election mechanism are explained in Section \ref{sec:the_gga_algorithm}. Section \ref{sec:experiment} presents experimental setups and result analyses. The last section concludes the paper. \rw{For better readability, an acronym table and a notation table have been presented in the Appendix section. }

\begin{figure*}[t]
    \centering
    \includegraphics[width=0.75\textwidth]{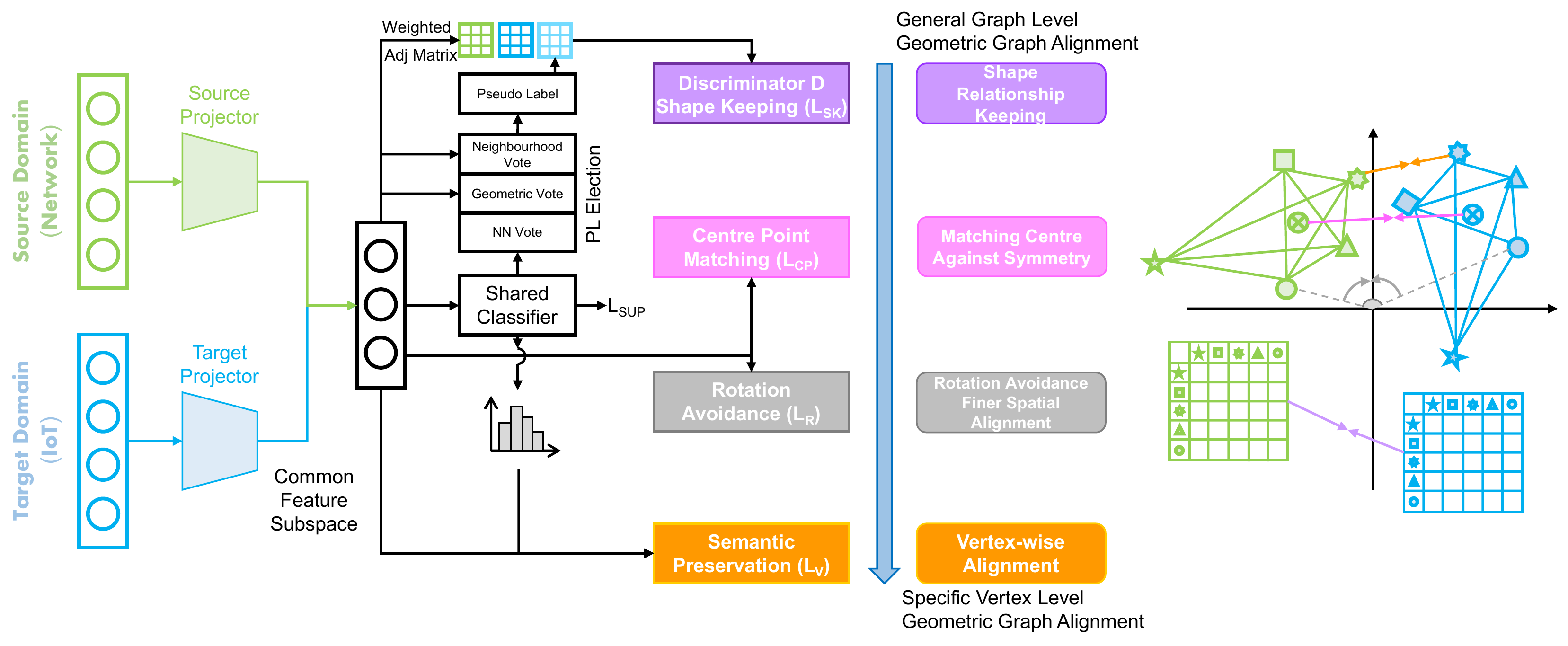}
    \caption{\rw{The architecture of the GGA method}}
    \label{fig:gga_architecture}
    \vspace{-0.5cm}
\end{figure*}

\section{Related Work}\label{sec:related_work}

\textit{IoT Intrusion Detection Methods} The IoT intrusion detection (IID) has drawn wide research attention to secure IoT infrastructures. Rule-based IID methods were initially popular. Dietz \cite{dietz2018iot} performed automatic scanning of neighbouring IoT devices for known vulnerability patterns and temporarily isolated detected compromised devices. Chen \cite{jun2014design} proposed to filter security violations via complex event processing, which required a sophisticated rule repository. On the other hand, machine learning techniques were also widely used for IID. Anthi \cite{anthi2019supervised} presented a three layer intrusion detector that worked under a supervised fashion for smart home settings. Shukla \cite{shukla2017ml} tackled the problem via a hybrid two-stage mechanism that combined Kmeans clustering and decision trees. On the deep learning perspective, models such as feedforward neural network, deep autoencoder and BiLSTM RNN were utilised to work on the IID problem by \cite{ge2019deep}, \cite{meidan2018n} and \cite{mcdermott2018botnet}, respectively. However, these methods either needed a sophisticated intrusion pattern repository, which requires substantial expertise to build and can hardly be complete and up-to-date, or required abundant amount of fully labelled data for training, which is labour-intensive to annotate. Hence, it naturally leads to domain adaptation-based methods which can comfortably work under the challenging data-scarce IID scenarios. 

\textit{Domain Adaptation and Intrusion Detection} Domain adaptation leverages source domain knowledge to facilitate better learning for data-scarce target domains, and hence is suitable to tackle the intrusion detection for the data-scarce IoT domain. Vu \cite{vu2020deep} utilised two autoencoders on source and target dataset and forced the alignment of the bottleneck layers. Later, Ning \cite{ning2021malware} leveraged the Laddernet to tackle IID under a semi-supervised setting. Hu \cite{hu2022deep} presented a deep subdomain adaptation network with attention mechanism that performed intrusion knowledge transfer by minimising local maximum mean discrepancy. However, these methods didn't use pseudo-label (PL) assignments to exploit unlabelled target domain, and failed to perform DA via a geometric graph-based approach, hence didn't preserve geometric properties during intrusion knowledge transfer. On the other hand, Chen \cite{chen2016transfer} tackled the intrusion domain alignment problem \rw{via Transfer Neural Tree} (TNT), a unified framework that combined feature mapping, adaptation and classification. A Generalised Joint Distribution Adaptation (G-JDA) approach was presented \cite{hsieh2016recognizing} \rw{to learn} a pair of feature projectors to eliminate the marginal and conditional distribution divergence. Yao \cite{yao2020discriminative} proposed the DDA method that applied an adaptive classifier to reduce distribution divergence and enlarge inter-class divergence. The TNT, G-JDA and DDA applied direct prediction as PLs for unlabelled target instances and completely ignored the label neighbourhood information. Yao \cite{yao2019heterogeneous} put forward the STN, a conditional distribution alignment strategy with the help of a soft-label paradigm. Singh \cite{singh2021improving} presented the STAR framework, which emphasised unlabelled target instances based on the distance with closest class prototypes during intrusion domain alignment. Saito \cite{saito2019semi} achieved intrusion knowledge transfer by optimising the minimax loss on the domain conditional entropy (MME). It utilised unlabelled target data based on a threshold-based strategy. The APE \cite{kim2020attract} method chose to alleviate intra-domain discrepancy via three procedures, namely Attraction, Perturbation and Exploration. From a clustering-based perspective, Li \cite{li2021cross} proposed Cross-Domain Adaptive Clustering (CDAC) to tackle the DA problem. In MME, APE and CDAC, unlabelled target instances will be pseudo-labelled based on a threshold strategy. However, when assigning PLs, these method either failed to jointly consider geometric properties, or assigned PLs in an isolated manner that ignored the relationships between pseudo-labelled instances and \rw{their} neighbouring labelled instances, compromising accurate intrusion knowledge transfer. Some methods also required a manual threshold set based on prior experience and \rw{was} not generalisable between tasks. 

Tackling intrusion domain alignment from a graph perspective is also feasible. For example, the WCGN method matched domains via graph learning \cite{wang2020prototype,3482115} to benefit the domain alignment. Pilanci \cite{pilanci2020domain} proposed a graph base alignment method by transferring the graph spectrum information. However, although these embedding-related methods can partially convey the geometric information of domains, learning a good embedding is highly data-dependent, hindering their applicability. Besides, none of these graph-based methods \rw{solved} the graph matching problem from a pure geometric graph alignment perspective, which left a void to be filled. 

Our method tackles the HDA problem from a pure geometric graph perspective, which jointly considers several levels of geometric property matching. The GGA method does not require a huge amount of data for graph embedding learning and enjoys a relatively low complexity. Besides, we utilise a pseudo-label election mechanism which jointly accounts for the geometric properties and the neighbourhood information, so that the confident but wrong PL prediction that violates geometric properties and near-boundary ambiguous PLs can be avoided for positive transfer. Finally, we utilise the GGA method to facilitate more accurate intrusion detection for the data-scarce IoT domain.

\section{Model Preliminary and Architecture}\label{sec:model_preliminary_and_architecture}

\subsection{Model Preliminary}\label{sec:model_preliminary}

The geometric graph alignment (GGA) method works under a semi-supervised heterogeneous DA setting. It involves a source NI domain formulated as follows: 
\begin{equation}\label{equ:source_ni_definition}
    \begin{split}
      & \mathcal{D}_{S} = \{\mathcal{X}_{S}, \mathcal{Y}_{N}\} = \{(x_{Si}, y_{Si})\}, i \in [1, n_{S}], \\
      & x_{Si} \in \mathbb{R}^{d_{S}}, y_{Si} \in [1, K]\,,
    \end{split}
\end{equation}
where the source NI domain contains $n_{S}$ traffic records with their corresponding intrusion label. Each record is represented using $d_{S}$ features, and there are $K$ categories. Similarly, the target II domain are defined as follows: 
\begin{equation}\label{equ:target_ii_definition}
    \begin{split}
      \mathcal{D}_{TL} &= \{\mathcal{X}_{TL}, \mathcal{Y}_{TL}\} = \{(x_{TLi}, y_{TLi})\}, i \in [1, n_{TL}], \\
      \mathcal{D}_{TU} &= \{\mathcal{X}_{TU}\} = \{(x_{TUj})\}, j \in [1, n_{TU}], \\
      \mathcal{D}_{T} &= \mathcal{D}_{TL} \cup \mathcal{D}_{TU}, x_{TLi}, x_{TUj} \in \mathbb{R}^{d_{T}}, y_{TLi} \in [1, K], \\
      n_{T} &= n_{TL} + n_{TU}, n_{TL} \ll n_{TU}\,.
    \end{split}
\end{equation}
Under the semi-supervised setting, the target domain is scarcely-labelled, i.e., $n_{TL} \ll n_{TU}$. The source and target domain present heterogeneities such as belonging to different feature spaces, i.e., $d_{S} \neq d_{T}$. 

\begin{figure*}[t]
    \centering
    \includegraphics[width=0.75\textwidth]{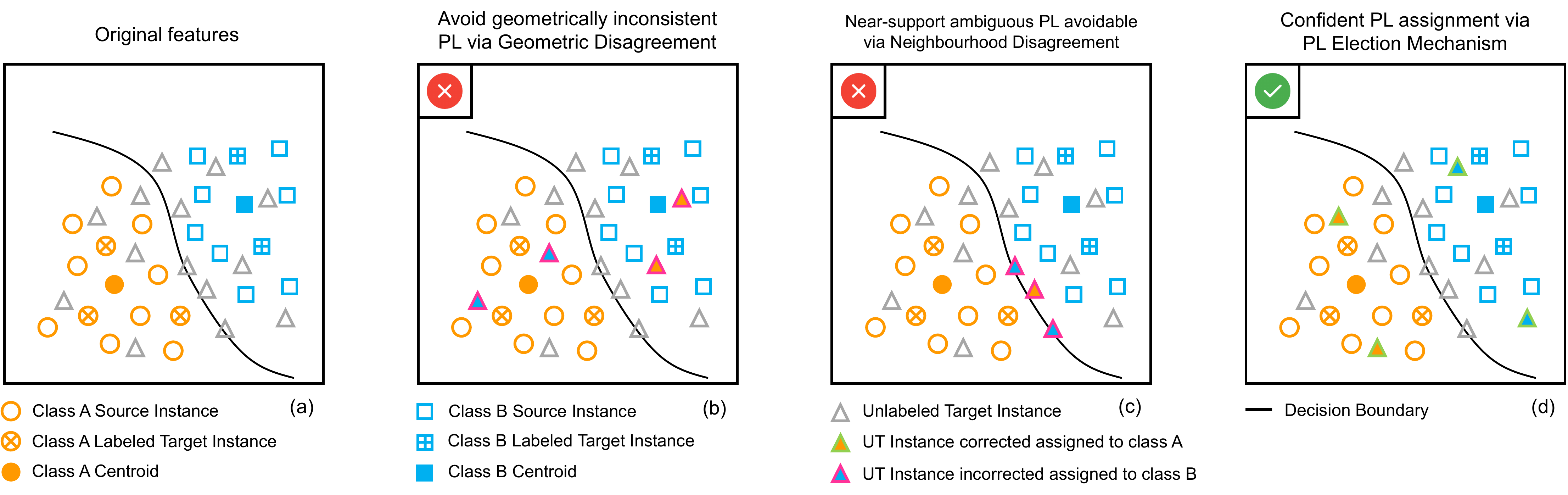}
    \caption{\rw{Illustration of the PLE. (a) the original feature; (b) avoid geometrically inconsistent PL via geometric disagreement; (c) near-boundary ambiguous PL assignment avoidable via neighbourhood disagreement; (d) PLE's assignment. }}
    \label{fig:pl_illustration}
    \vspace{-0.5cm}
\end{figure*}

\subsection{Graph Formulation}\label{sec:graph_formulation}

To perform the geometric graph alignment, we formulate each intrusion domain as a graph, \rw{i.e., $G_{X} = <V_{X}, E_{X}>$, $X \in \{S, T\}$. }Both domains share $K$ intrusion categories, therefore, each graph has $K$ vertices, the vertex $V_{i}$ is the centroid of the category $i$, denoted as follows: 
\rw{\begin{equation}
    V_{X}^i = \frac{1}{n_{X}^i} \sum_{j=1}^{n_{X}^i} x_{X_j}, i \in [1, K], X \in \{S, T\}\,,
\end{equation}}
where $n^i$ is the number of records under category $i$. Both graphs are formulated as a complete graph, the weight of edge $E<V_{X}^i, V_{X}^j>$ is \rw{set to be} the Euclidean distance between vertex $V_{X}^i$ and $V_{X}^j$. The corresponding weighted adjacency matrices (WAMs) are denoted as $M_{S}$ and $M_{T}$.

\subsection{Model Architecture and Overview}\label{sec:model_architecture}

The architecture of the GGA method has been shown in Fig. \ref{fig:gga_architecture}. Each intrusion domain has a feature projector that maps features into a common feature subspace with dimension $d_C$. The feature projector is defined as follows: 
\begin{equation}
    \begin{split}
        & f(x_i) = \begin{cases}
            E_{S}(x_i) & \text{if $x_i \in \mathcal{X}_{S}$} \\
            E_{T}(x_i) & \text{if $x_i \in \mathcal{X}_{T} = \mathcal{X}_{TL} \cup \mathcal{X}_{TU}$}
          \end{cases} \\
        & f(x_i) \in \mathbb{R}^{d_C}\,.
    \end{split}
\end{equation}
The GGA method will then utilise the pseudo-label election (PLE) mechanism to assign fine-grained PLs to unlabelled target data and avoid error-prone PLs to mitigate negative transfer. To perform geometric graph alignment between these heterogeneous domains, the WAM of the source data, the labelled target data, and the combination of labelled and pseudo-labelled target data will be generated to confuse the discriminator $D$. Highly similar WAMs indicate well-aligned intrusion categories and the fine preservation of category-wise interrelations, and is equivalent with a perfect geometric graph shape keeping. By fusing three WAMs, the geometric shape of domain graphs are aligned, meanwhile the labelled and pseudo-labelled target data will be better fused together. After keeping the shape, the domain graphs can still misalign due to rotation and symmetry, which are mitigated by the rotation avoidance mechanism and the centre point matching. Besides, categorical correlations yielded by the shared classifier $C$ will be preserved between domains, which acts as a vertex-level alignment mechanism. Holistically, the GGA approach aligns the domain graphs from general shape level to specific vertex level. The motivation is to align the domain graphs in a fine-grained manner, such that the shared classifier $C$ yields the best intrusion detection accuracy for unlabelled target domain.

\section{The GGA Algorithm}\label{sec:the_gga_algorithm}

In this section, we will firstly introduce the pseudo-label election mechanism which facilitates better target participation during the geometric graph alignment process. Then, the geometric graph alignment process and its constituting components will be explained.

\subsection{Pseudo-label Election Mechanism}\label{sec:pseudo_label_election_mechanism}

Assigning pseudo-labels to unlabelled target data can excavate its potentials during intrusion knowledge transfer. However, erroneous PL assignment may mislead the model towards negative transfer. Traditional efforts mainly assigned PL to instances in an isolated manner, without considering the relationship between labels, and suffered from issues such as confident but \rw{geometrically-inconsistent} PLs and near-support ambiguous PLs. Therefore, we utilise the Pseudo-Label Election (PLE) mechanism to mitigate these issues as much as possible. The PLE jointly considers the voting of NN prediction, the geometric properties and neighbourhood information. A PL assignment will only be made for an instance if these three votes reach a consensus. When producing the geometric property-based PL, the category of the most Cosine-similar labelled data centroid $\mu_{S+TL}^i$ will be utilised as $PL_G$ for each unlabelled target instance and is defined as follows: 
\begin{equation}
    \begin{split}
        PL_{G}^{i} &= \argmax_k COS(\mu_{S+TL}^{(k)}, x_{TU}^i), \\
        \mu_{S+TL}^{(k)} &= \frac{1}{n_{S}^{(k)} + n_{TL}^{(k)}} \sum_{x_i \in \mathcal{X}_{S}^{(k)} \cup \mathcal{X}_{TL}^{(k)}} x_i\,,
    \end{split}
\end{equation}
where $PL_{G}^i$ denotes the geometric-based PL for the $x_{TU}^{i}$, $\mathcal{X}_{S}^{(k)}$ denotes source instances from category $k$. \rw{If the NN-prediction yields a confident PL prediction but is inconsistent with the geometric similarity property, then such confident but contradictary PL will be rejected as illustrated in Fig. \ref{fig:pl_illustration} (b)}. Besides, the PLE will also consult the neighbourhood information when assigning PLs. If the K-nearest neighbourhood around an unlabelled target instance cannot reach a majority agreement, or reach an agreement against the NN prediction or the geometric vote, then such assignment will also be rejected. This is useful especially when deciding the PL for near-boundary unlabelled instances, as illustrated in Fig. \ref{fig:pl_illustration} (c). Since the neighbourhood can be harder to reach an agreement near the boundary due to ambiguity, the PLE can effectively get rid of near-boundary PLs \rw{which are} more likely to be erroneous. Overall, the PLE will only assign confident PLs with probabilistic, geometric and neighbourhood soundness, which can significantly boost the PL accuracy and therefore lead to positive intrusion knowledge transfer.

\subsection{Geometric Graph Alignment}\label{sec:geometric_graph_alignment}

\begin{figure*}[t]
    \centering
    \includegraphics[width=0.75\textwidth]{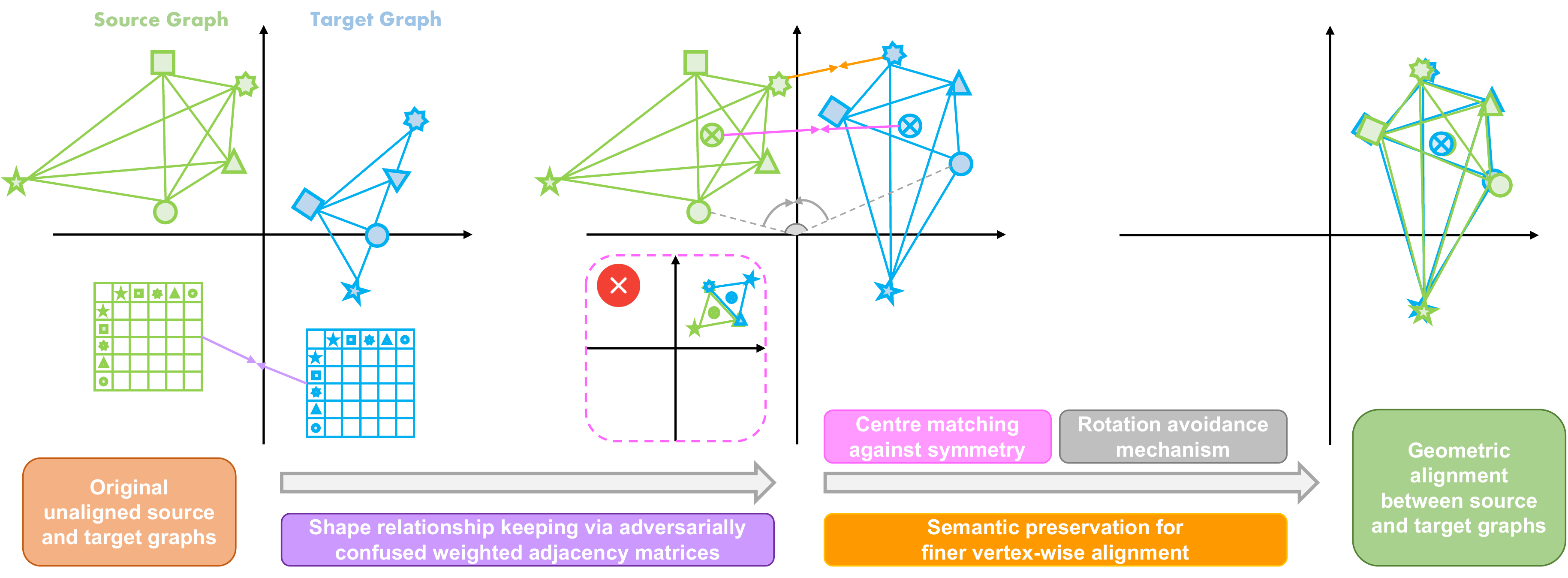}
    \caption{Illustration of the Geometric Graph Alignment (GGA) method. }
    \label{fig:gga_illustration}
    \vspace{-0.5cm}
\end{figure*}

The GGA method has been illustrated in Fig. \ref{fig:gga_illustration}. It will perform geometric graph alignment from the general graph granularity to the specific vertex granularity, i.e., the shape keeping via confused discriminator (\textit{purple step}), rotation avoidance mechanism (\textit{grey step}), centre point matching against symmetry (\textit{pink step}) and the vertex-level alignment via semantic preservation (\textit{orange step}). 

\textbf{Shape Keeping} Firstly, the GGA with align the graph shape via a confused discriminator. We define the \textit{Same Shape Rule} as follows: 
\begin{definition}\label{def:same_shape_rule}
    \textbf{Same Shape Rule}: Both $G_{S}$ and $G_{T}$ have their shape aligned with each other if and only if the weighted adjacency matrices (WAMs) $M_{S}$ and $M_{T}$ are the same. 
\end{definition}
Specifically, with the help from the PLE, the GGA will construct three WAMs: the source WAM $M_{S}$, the labelled target WAM $M_{TL}$ and a WAM based on both labelled and pseudo-labelled target data $M_{TL+PL}$. \rw{These WAMs will then be flattened and feed into the discriminator $D$, a single-layer neural network that tries to distinguish the origin of the input WAM. The source domain WAM $M_{S}$ is assigned with domain label 1, while target domain WAMs are assigned with domain label 0.} The shape keeping loss $\mathcal{L}_{SK}$ is defined as follows: 
\begin{equation}
    \begin{split}
        \mathcal{L}_{SK} &= \log(D(M_{S}))\\
        &+ \frac{1}{2} \sum_{M \in \{M_{TL}, M_{TL+PL}\}} (1 - \log(D(M)))
    \end{split}
\end{equation}
\rw{The source and target projector will try to minimise the $\mathcal{L}_{SK}$ and let the discriminator $D$ to be unable to distinguish the origin of the input WAMs, while the discriminator will try to stay unconfused. }Upon this minimax game reaches an equilibrium, both $G_{S}$ and $G_{T}$ will have their shape aligned as indicated in Fig. \ref{fig:gga_illustration} (Mid), and the labelled and pseudo-labelled target data will be better fused together. 

\textbf{Rotation Avoidance against Rotated Misalignment} As indicated in Fig. \ref{fig:gga_illustration} (Mid), graph rotation can still cause domain graph misalignment, even though the shape is aligned. Therefore, to further align the domain graphs geometrically, we define the \textit{Same Angle Rule} as follows: 
\begin{definition}\label{def:same_angle_rule}
    \textbf{Same Angle Rule} For graph $G_{S}$ and $G_{T}$, the Same Angle Rule holds if $\forall i \in [1, K], 1 - COS(V_{S}^i, V_{T}^i) = 0$. 
\end{definition}
The GGA method will keep the Same Angle Rule holding by minimising the rotation loss, defined as follows: 
\begin{equation}
    \mathcal{L}_{R} = \sum_{i=1}^{K} (1 - COS(V_{S}^i, V_{T}^i))
\end{equation}
By minimising $\mathcal{L}_{R}$, graph misalignment caused by rotation will be prevented since the categorise-wise vertex angle is enforced to be $0^{\circ}$. 

\textbf{Centre Point Matching against Symmetric Misalignment} As shown in the pink-boxed example in Fig. \ref{fig:gga_illustration} (Mid), upon fixing the graph shape and enforcing the Same Angle Rule, symmetry can still cause misalignment. Therefore, we further define the \textit{Same Centre Rule} as follows: 
\rw{\begin{definition}
    \textbf{Same Centre Rule} For graph $G_{S}$ and $G_{T}$, the Same Centre Rule holds if $\mu_{S} = \mu_{T}$, $\mu_{X} = \frac{1}{n_{X}} \sum_{i=1}^{n_{X}} {\mathcal{X}_{X}}_{i}, X \in \{S, T\}$\,.
\end{definition}}
We minimise the \rw{centre point matching loss} as follows: 
\begin{equation}
    \mathcal{L}_{CP} = ||\mu_{S} - \mu_{T}||^2
\end{equation}
Enforcing the Same Centre Rule brings three-fold advantages: Firstly, the graph misalignment caused by symmetry can be prevented; Then, it boosts data participation during the intrusion knowledge transfer to fully excavate the potentials in all data, irrespective of whether a PL is assigned or not; Finally, despite its computational simplicity, it can boost the intrusion detection performance as indicated by the experiments. 

\textbf{Vertex-level Alignment via Semantic Preservation} The above three steps focus on the overall graph level, in this step, we shift our focus to the vertex granularity. Each vertex in the domain graph represents an intrusion category centroid, during prediction, it presents a unique probabilistic category-wise correlation. Use object as an example, a laptop should be highly similar with other laptops, somewhat similar with TV screen, and very dissimilar with a bike, irrespective of its domain origin. By enforcing the corresponding vertices from both domain graphs to preserve the correlation semantic, it in turn forces vertex-level alignment between domain graphs. Specifically, for source category $k$, its correlation semantic is defined as follows: 
\begin{equation}
    q^{(k)} = \frac{1}{n_{S}^{(k)}} \sum_{x_i \in \mathcal{X}_{S}^{(k)}} \text{softmax}(\frac{C(f(x_i))}{T})\,,
\end{equation}
where $\mathcal{X}_{S}^{(i)}$ denotes category $i$ source instances, $C$ and $f$ denote the shared classifier and the feature projector, respectively, $T$ is a temperature hyperparameter that controls the correlation semantic smoothness. Similarly, the correlation semantic of each labelled target instance is defined as follows: 
\begin{equation}
    p_{i} = \text{softmax}(C(f(x_i))), x_i \in \mathcal{X}_{TL}
\end{equation}
To perform the semantic preservation, we minimise the vertex-level alignment loss as follows: 
\begin{equation}
    \mathcal{L}_{VS} = - \frac{1}{n_{TL}} \sum_{(x_i, y_i) \in (\mathcal{X}_{TL}, \mathcal{Y}_{TL})} (q^{(y_i)})^{\top} \log(p_i)
\end{equation}
Together with the supervision provided by the labelled target instances, the final vertex-level alignment loss is defined as follows: 
\begin{equation}
    \begin{split}
        \mathcal{L}_{V} &= \frac{1 - \alpha}{n_{TL}} \sum_{(x_i, y_i) \in (\mathcal{X}_{TL}, \mathcal{Y}_{TL})} \mathcal{L}_{ce}(C(f(x_i)), y_i)\\
        &+ \alpha \mathcal{L}_{VS}\,,
    \end{split}
\end{equation}
where $\mathcal{L}_{ce}$ denotes cross entropy loss. By minimising the vertex-level alignment loss $\mathcal{L}_{V}$, it will enforce vertices in the same category from different graphs to align with each other. Otherwise, the correlation semantic will fail to be preserved. 

\textbf{Geometric Graph Alignment Theorem} The GGA forms the above mechanism into a holistic framework, and can align two domain graphs with theoretical guarantee. We state the \textit{GGA Theorem} as follows: 
\begin{theorem}\label{the:gga_theorem}
    \textbf{GGA Theorem} Given graph $G_{S}$ and $G_{T}$, if the Same Shape Rule, the Same Angle Rule and the Same Centre Rule hold simultaneously, then $G_{S}$ and $G_{T}$ must exactly align with each other. 
\end{theorem}
The proof of the GGA Theorem is as follows: 
\begin{proof}
    We prove the GGA Theorem by \textit{induction with the help of contradiction}. 

    Case 1: Both $G_{S}$ and $G_{T}$ have $2$ vertices. Given that the Same Angle Rule holds, both graphs must be parallel with each other. Given that the Same Shape Rule holds, it enforces the only edge in both graphs to have equal length. Therefore, it is trivial to conclude that these two graphs are the same. 

    Case k: Both $G_{S}$ and $G_{T}$ have $k$ vertices. The aforementioned $3$ rules hold and $G_{S}$ and $G_{T}$ are aligned. We add an additional vertex to $G_{S}$ and $G_{T}$ separately. Without breaking any of the aforementioned $3$ rules, the new graph $G_{S}'$ and $G_{T}'$ also align with each other. 

    \textit{Proof Case k by contradiction: }
    
    Since under Case k, the prerequisite states that $G_{S}$ and $G_{T}$ align with each other, therefore, we simply denote both of them as $G_{X}$. 

    Case k1: If $G_{X}$ is asymmetric regarding any line, then it is trivial that there is no possible strategy to add point differently to $G_{X}$ to get $G_{X}'$ and $G_{X}''$ without violating the Same Centre Rule. 

    Case k2. If $G_{X}$ is symmetric regarding a symmetric axis, to stick to the Same Shape Rule and the Same Angle Rule, the only possible strategy to add $V_{S}^{k+1}$ and $V_{T}^{k+1}$ is as follows: the line crossing $V_{S}^{k+1}$ and $V_{T}^{k+1}$ should also cross origin (Same Angle Rule) and the centre point of the symmetric axis (Same Shape Rule). However, if we add $V_{S}^{k+1}$ and $V_{T}^{k+1}$ differently, then they must be symmetric regarding the graph symmetric axis, which violates the Same Centre Rule for the graphs. 
\end{proof}

Given that the GGA theorem holds, the constituting components of the GGA method can achieve a fine-grained geometric graph alignment with theoretical guarantee and benefit intrusion knowledge transfer.

\subsection{Overall Optimisation Objective}\label{sec:overall_optimisation_objective}

Finally, the source labels provide supervision during the training process with supervision loss defined as follows: 
\begin{equation}
    \mathcal{L}_{SUP}(\mathcal{X}_{S}, \mathcal{Y}_{S}) = \frac{1}{n_{S}} \sum_{(x_i, y_i) \in (\mathcal{\mathcal{X}_{S}, \mathcal{Y}_{S}})} \mathcal{L}_{ce} (c(f(x_i)), y_i)\,.
\end{equation}
Overall, the optimisation objective of GGA is as follows: 
\begin{equation}\label{equ:overall_objective}
    \min_{C, E_S, E_T} \max_{D} (\mathcal{L}_{SUP} + \gamma \mathcal{L}_{SK} + \eta \mathcal{L}_{R} + \lambda \mathcal{L}_{CP} + \mathcal{L}_{V})\,,
\end{equation}
where $\gamma$, $\eta$ and $\lambda$ are hyperparameters controlling the influence of loss components during optimisation. During initial training stages, both domain graphs may suffer from immature shape, therefore the $\gamma$ is set to a relatively low value to emphasise other components such as vertex-wise semantic alignment, etc. \rw{As the training progresses}, the $\gamma$ will grow linearly to gradually emphasise the importance of shape keeping. \rw{To form the optimisation into an end-to-end procedure, we follow \cite{ganin2015unsupervised} to apply the gradient reversal layer for the discriminator and optimise the model using Adam gradient descent. Upon the equilibrium of the above minimax game is reached, }the network training concludes, and the domain graphs can be aligned in a fine-grained manner, which facilitates \rw{more accurate intrusion detection} for the target IoT domain.

\section{Experiment}\label{sec:experiment}

{\renewcommand{\arraystretch}{1.2}%
\begin{table*}[t]
    \centering
    \caption{Intrusion detection accuracy under default data scarcity ratio $n_{TL}:n_{TU} = 1:50$}
    \label{tab:grand_result_table}
    \begin{tabular}{c|ccccccccc|c}
    \Xhline{2\arrayrulewidth}
    S $\rightarrow$ T & C $\rightarrow$ B & N $\rightarrow$ B & K $\rightarrow$ B & C $\rightarrow$ G & N $\rightarrow$ G & K $\rightarrow$ G & C $\rightarrow$ W & N $\rightarrow$ W & K $\rightarrow$ W & Avg \\ \hline
    TNT & 10.71 & 16.42 & 16.40 & 53.85 & 70.07 & 61.37 & 53.41 & 70.07 & 61.36 & 45.96 \\
    APE & 33.99 & 51.46 & 52.42 & 54.18 & 70.35 & 61.67 & 47.24 & 63.35 & 55.81 & 54.50 \\
    MME & 34.98 & 51.35 & 52.67 & 54.14 & 70.29 & 61.63 & 50.44 & 65.46 & 57.81 & 55.42 \\
    CDAC & 34.04 & 50.29 & 50.87 & 54.22 & 70.35 & 61.70 & 53.89 & 70.34 & 61.65 & 56.37 \\
    STAR & 33.69 & 50.87 & 51.02 & 54.17 & 70.32 & 61.65 & 53.72 & 70.23 & 61.83 & 56.39 \\
    DDAS & 35.13 & 52.15 & 52.01 & 53.90 & 72.52 & 60.11 & 54.35 & 71.54 & 63.48 & 57.24 \\
    STN & 35.93 & 51.94 & 53.48 & 57.44 & 69.85 & 64.84 & 54.28 & 71.28 & 63.24 & 58.03 \\
    DDAC & 35.20 & 52.54 & 53.55 & 55.65 & 71.92 & 59.36 & 56.96 & 71.80 & 65.34 & 58.04 \\
    WCGN & 42.79 & 60.10 & 63.80 & 58.26 & 76.62 & 66.83 & 56.82 & 72.39 & 64.41 & 62.45 \\ \hline
    \rowcolor{gray}
    \textbf{GGA (Ours)} & \textbf{48.59} & \textbf{68.99} & \textbf{77.26} & \textbf{59.43} & \textbf{79.18} & \textbf{68.25} & \textbf{58.71} & \textbf{72.75} & \textbf{66.22} & \textbf{66.60} \\ \Xhline{2\arrayrulewidth}
    \end{tabular}
\end{table*}}

{\renewcommand{\arraystretch}{1.2}%
\begin{table*}[t]
    \centering
    \caption{Intrusion detection accuracy under varied $n_{TL}:n_{TU}$ ratios}
    \label{tab:ratio_performance_table}
    \begin{tabular}{c|ccccccccc|cc}
    \Xhline{2\arrayrulewidth}
    S $\rightarrow$ T & \multicolumn{3}{p{3.4cm}}{\centering N $\rightarrow$ B} & \multicolumn{3}{p{3.4cm}}{\centering K $\rightarrow$ G} & \multicolumn{3}{p{3.4cm}|}{\centering C $\rightarrow$ W} & Overall & $1:100$ \\ \cline{1-10}
    $n_{TL}:n_{TU}$ & $1:10$ & $1:50$ & $1:100$ & $1:10$ & $1:50$ & $1:100$ & $1:10$ & $1:50$ & $1:100$ & Avg & Avg \\ \hline
    TNT & 18.86 & 16.42 & 16.38 & 61.40 & 61.37 & 61.30 & 53.44 & 53.41 & 53.31 & 43.99 & 43.66 \\
    APE & 51.84 & 51.46 & 47.14 & 61.70 & 61.67 & 61.50 & 53.75 & 47.24 & 31.76 & 52.01 & 46.80 \\
    MME & 52.84 & 51.35 & 47.01 & 61.65 & 61.63 & 45.23 & 53.36 & 50.44 & 50.07 & 52.62 & 47.44 \\
    CDAC & 50.70 & 50.29 & 50.03 & 61.73 & 61.70 & 61.32 & 54.24 & 53.89 & 53.64 & 55.28 & 55.00 \\
    STAR & 51.96 & 50.87 & 50.75 & 61.70 & 61.65 & 61.40 & 53.75 & 53.72 & 53.70 & 55.50 & 55.28 \\
    DDAS & 52.50 & 52.15 & 51.64 & 60.37 & 60.11 & 59.70 & 54.56 & 54.35 & 53.71 & 55.45 & 55.02 \\
    STN & 53.71 & 51.94 & 51.65 & 65.15 & 64.84 & 61.03 & 58.68 & 54.28 & 54.05 & 57.26 & 55.58 \\
    DDAC & 53.16 & 52.54 & 52.25 & 59.93 & 59.36 & 58.61 & 58.17 & 56.96 & 56.60 & 56.40 & 55.82 \\
    WCGN & 61.98 & 60.10 & 58.47 & 67.09 & 66.83 & 65.92 & 58.37 & 56.82 & 55.17 & 61.19 & 59.85 \\ \hline
    \rowcolor{gray}
    \textbf{GGA (Ours)} & \textbf{72.87} & \textbf{68.99} & \textbf{68.26} & \textbf{68.73} & \textbf{68.25} & \textbf{67.79} & \textbf{59.04} & \textbf{58.71} & \textbf{57.30} & \textbf{65.55} & \textbf{64.45} \\ \Xhline{2\arrayrulewidth}
    \end{tabular}
\end{table*}}

\subsection{Experimental Datasets}\label{sec:experimental_datasets}

During experiments, we utilise $5$ representative and comprehensive intrusion detection datasets, \rw{which include} $3$ network intrusion datasets: NSL-KDD, UNSW-NB15 and CICIDS2017, and $2$ IoT intrusion datasets: UNSW-BOTIOT and UNSW-TONIOT. 

\textbf{Network Intrusion Dataset: NSL-KDD} The NSL-KDD dataset \cite{tavallaee2009detailed} was released in 2009, which addressed issues of the prior dataset KDD CUP99 \cite{hettich1999kdd} such as having lots of redundant records. The NSL-KDD dataset contains benign traffic with $4$ types of intrusions, such as probing attack, denial of service (DoS) attack, etc. Follow \cite{anthi2019supervised}, we reasonably utilise $20\%$ of the dataset. Each traffic record in the dataset is represented using $41$ features. We follow Harb \cite{harb2011selecting} to use the top $31$ most informative features as the feature representation. The dataset is denoted as $K$. 

\textbf{Network Intrusion Dataset: UNSW-NB15} The UNSW-NB15 dataset \cite{moustafa2015unsw} was created by UNSW in 2015 using the IXIA PerfectStorm tool, which aimed to address the data quality issue and out-of-date incomprehensive network flow issue observed in previous datasets. The dataset contains $10$ traffic categories, including normal traffic, DoS attacks, reconnaissance attacks, etc. We utilise $2700$ traffic records, which follows the dataset magnitude in \cite{alkadi2020deep}. The dataset is represented using $49$ features, we perform preprocessing to remove $4$ features having value $0$ for nearly all records. The dataset is denoted as $N$. 

\textbf{Network Intrusion Dataset: CICIDS2017} The CICIDS2017 dataset \cite{sharafaldin2018toward} was released in 2017, which served as one of the most up-to-date network intrusion \rw{datasets} with modern attack patterns. The dataset has 7 types of intrusions with benign traffic, represented in $77$-dimensional features. We utilise the $20\%$ portion of the dataset provided by the dataset creator to perform the model training and testing. During preprocessing, we perform data deduplication and categorical-numerical entries conversion. Following Stiawan \cite{stiawan2020cicids}, we use features with top $40$ information gain as the feature space of the dataset and denote the dataset as $C$. 

\textbf{IoT Intrusion Dataset: UNSW-BOTIOT} The UNSW-BOTIOT dataset \cite{koroniotis2019towards} was released in 2017 by UNSW, which presented up-to-date modern attack scenarios captured based on a realistic testbed environment. The testbed environment deployed IoT devices such as weather station, smart fridge, etc., and utilised MQTT protocol, a lightweight IoT communication protocol commonly used in realistic IoT scenarios. The dataset quality has been carefully addressed, and the attack diversity has been improved. The dataset contains $4$ categories including normal traffic, DoS attacks, information theft attacks, etc. Following \cite{alkadi2020deep}, we utilise $10000$ data records. The original dataset utilises a $46$-dimensional feature space. We follow the dataset creator's advice to use top $10$ most informative features as the feature space. The dataset is denoted as $B$. 

{\renewcommand{\arraystretch}{1.2}%
\begin{table*}[t]
    \centering
    \caption{Intrusion detection performance using various evaluation metrics under default data scarcity}
    \label{tab:metrics_performance_table}
    \begin{tabular}{c|cccccccc}
    \Xhline{2\arrayrulewidth}
    S $\rightarrow$ T & \multicolumn{4}{p{4cm}}{\centering N $\rightarrow$ G} & \multicolumn{4}{p{4cm}}{\centering K $\rightarrow$ B} \\ \hline
    Metrics & P & R & F & A & P & R & F & A \\ \hline
    APE & 0.495 & 0.703 & 0.581 & 0.500 & 0.395 & 0.516 & 0.366 & 0.557 \\
    CDAC & 0.493 & 0.702 & 0.579 & 0.500 & 0.251 & 0.501 & 0.335 & 0.553 \\
    STAR & 0.495 & 0.703 & 0.581 & 0.491 & 0.250 & 0.500 & 0.333 & 0.481 \\
    DDAS & 0.700 & 0.718 & 0.688 & 0.696 & 0.390 & 0.522 & 0.376 & 0.674 \\
    STN & 0.495 & 0.703 & 0.581 & 0.693 & 0.525 & 0.534 & 0.405 & 0.768 \\
    DDAC & 0.749 & 0.712 & 0.713 & 0.700 & 0.393 & 0.540 & 0.407 & 0.769 \\ 
    WCGN & 0.763 & 0.747 & 0.740 & 0.884 & 0.608 & 0.617 & 0.583 & 0.743 \\ \hline
    \rowcolor{gray}
    \textbf{GGA (Ours)} & \textbf{0.828} & \textbf{0.800} & \textbf{0.790} & \textbf{0.927} & \textbf{0.778} & \textbf{0.773} & \textbf{0.757} & \textbf{0.884} \\ \Xhline{2\arrayrulewidth}
    \end{tabular}
\end{table*}}

\textbf{IoT Intrusion Dataset: UNSW-TONIOT} The UNSW-TONIOT dataset \cite{booij2021ton_iot} serves as one of the latest IoT intrusion datasets \cite{abdelmoumin2021performance}, released in 2021. It reflects modern IoT standards, protocols, and is captured on modern testbed consists of $7$ types of IoT devices, such as smart fridge, modbus sensor, GPS tracker, etc. The dataset covers $9$ types of intrusions, including scanning attacks, DoS attacks, etc. Heterogeneities present between IoT devices as \rw{the features} captured by each type of \rw{IoT device} have their own feature dimension. Following \cite{alkadi2020deep,qiu2020adversarial}, we utilise around $10\%$ of the dataset, and select the \rw{weather meter} and GPS tracker as the IoT devices used during experiments, which are denoted as $W$ and $G$, respectively. 

\textbf{Comprehensiveness of Datasets} The datasets we utilised are representative and comprehensive to verify the effectiveness of the proposed method. Firstly, these datasets are widely recognised and widely \rw{adopted by} the research community to testify intrusion detection effectiveness with thousands of citations. Secondly, these datasets are developed and released in recent years, some of them are release in 2021, \rw{therefore, }they can reflect current intrusion trends and methods. Finally, these datasets are captured on realistic testbeds with large-scale real IoT devices, and the sufficiency of the testbed is recognised by the research community. Hence, these datasets are representative with guaranteed comprehensiveness.

\subsection{Shared Intrusion}\label{sec:shared_intrusion}

The network intrusion datasets and the IoT intrusion datasets can have at most $8$ shared categories that can be transferred as intrusion knowledge, such as DoS attack, password attack, backdoor attack, etc. These shared intrusion categories account for $100\%$, $54.9\%$, $100\%$, $100\%$ and $98.3\%$ amount of records in NSL-KDD, UNSW-NB15, CICIDS2017, UNSW-BOTIOT and UNSW-TONIOT dataset, respectively. Therefore, transferring intrusion knowledge with wide coverage can effectively detect most modern intrusions targeting the IoT domain.

\subsection{Implementation Details}\label{sec:implementation_details}

We implement the GGA method using the PyTorch deep learning framework. Following \cite{li2020simultaneous,yao2019heterogeneous}, feature projectors are two-layer neural networks using LeakyReLu \cite{maas2013rectifier} as their activation function. Both the shared classifier and the discriminator are implemented as single-layer neural networks. The hyperparameter setting is fixed during all experiments as follows: $\alpha = 0.1$, $\gamma_{min} = 0.01$, $\gamma_{max} = 0.1$, $\eta = 0.01$, $\lambda = 0.01$, $d_{C} = 3$, $T = 5$ and $\#neighbour = 4$. Note that $\gamma$ will increase linearly from $\gamma_{min}$ to $\gamma_{max}$ as the training progresses. To emphasise the target data scarcity, we set $n_{TL}:n_{TU} = 1:50$ as the default ratio. We also conduct the parameter sensitivity analyses to verify the stable and robust performance of the GGA method. Following \cite{alkadi2020deep,li2018ai}, we use unlabelled target prediction accuracy as our major evaluation metric, and also use the category-weighted precision (P), recall (R), F1-score (F) and Area under the ROC Curve (A) \cite{ferrag2020deep,zavrak2020anomaly} to evaluate the GGA performance. \rw{Specifically, we define true positive $TP^{(k)}$ as the number of category $k$ intrusions being corrected identified, similar for true negative $TN^{(k)}$, false positive $FP^{(k)}$ and false negative $FN^{(k)}$. The mathematical definitions of evaluation metrics are as follows: }
\rw{\begin{equation}
\begin{split}
    Accuracy & = \frac{\sum_{k=1}^{K} (TP^{(k)} + TN^{(k)})}{n_{TU}}\,,
\end{split}
\end{equation}}
\rw{\begin{equation}
\begin{split}
    Precision & = \sum_{k=1}^{K} \frac{|\mathcal{X}_{TU}^{(k)}|}{n_{TU}} \cdot \frac{TP^{(k)}}{TP^{(k)} + FP^{(k)}}\\
    & = \sum_{k=1}^{K} \frac{|\mathcal{X}_{TU}^{(k)}|}{n_{TU}} \cdot Precision^{(k)}\,,
\end{split}
\end{equation}}
\rw{\begin{equation}
\begin{split}
    Recall & = \sum_{k=1}^{K} \frac{|\mathcal{X}_{TU}^{(k)}|}{n_{TU}} \cdot \frac{TP^{(k)}}{TP^{(k)} + FN^{(k)}}\\
    & = \sum_{k=1}^{K} \frac{|\mathcal{X}_{TU}^{(k)}|}{n_{TU}} \cdot Recall^{(k)}\,,
\end{split}
\end{equation}}
\begin{equation}
    \rw{F1 = \sum_{k=1}^{K} \frac{|\mathcal{X}_{TU}^{(k)}|}{n_{TU}} \cdot \frac{2 \cdot Precision^{(k)} \cdot Recall^{(k)}}{Precision^{(k)} + Recall^{(k)}}\,.}
\end{equation}
\rw{Besides, metrics A (AUC) represents the area under the ROC curve, a curve plotting the TP rate and the FP rate. }

\subsection{State-of-the-art Baselines}\label{sec:state_of_the_art_baselines}

We utilise $9$ state-of-the-art comparing methods, including TNT \cite{chen2016transfer}, MME \cite{saito2019semi}, STN \cite{yao2019heterogeneous}, APE \cite{kim2020attract}, DDAS, DDAC \cite{yao2020discriminative}, WCGN \cite{wang2020prototype,3482115}, CDAC \cite{li2021cross} and STAR \cite{singh2021improving}. All of them are from top-tier conferences and journals, and $8$ of them are proposed between $2019$ and $2021$. We summarise their differences with GGA as follows: 
\begin{itemize}
    \item From the pseudo labelling perspective, the DDAC, DDAS, WCGN and TNT utilise predicted hard pseudo label for target instances and ignore both the geometric property and neighbourhood information. On the other hand, MME, CDAC and APE involve threshold-based pseudo label strategy. However, setting thresholds properly requires expertise experience and has compromised flexibility. Both APE and STAR apply pseudo label strategy using geometric property as reference, however, the neighbourhood information is still ignored. Besides, STN utilises a soft-label strategy, lacking emphasise on confident predictions. The GGA is the only method that jointly considers both the geometric property and neighbourhood information, while avoiding hard-to-set threshold. 
    \item From the domain alignment perspective, these baselines apply diverse techniques such as CDAC's adversarial adaptive clustering, MME's alternated conditional entropy minimisation, STN's joint distribution matching, etc. However, these methods fail to explicitly conduct domain alignment from a geometric graph perspective. To our best knowledge, there lacks a similar pure geometric-based graph baseline method, the WCGN is a comparable state-of-the-art graph method based on graph learning framework. However, since a proper graph learning requires both sufficient data and a high complexity, it is challenging under the data-scarce and computationally-constrained IoT scenario. \rw{Conversely, }the GGA performs knowledge transfer via a geometric graph alignment perspective. It fills the void of previous methods, \rw{avoids }heavy data dependency, and has a relatively low complexity. 
\end{itemize}
Therefore, these state-of-the-art baseline methods are representative and can be used to verify the superiority of the GGA method from different perspectives.

\begin{figure}[!ht]
    \centering
    \includegraphics[width=0.97\columnwidth]{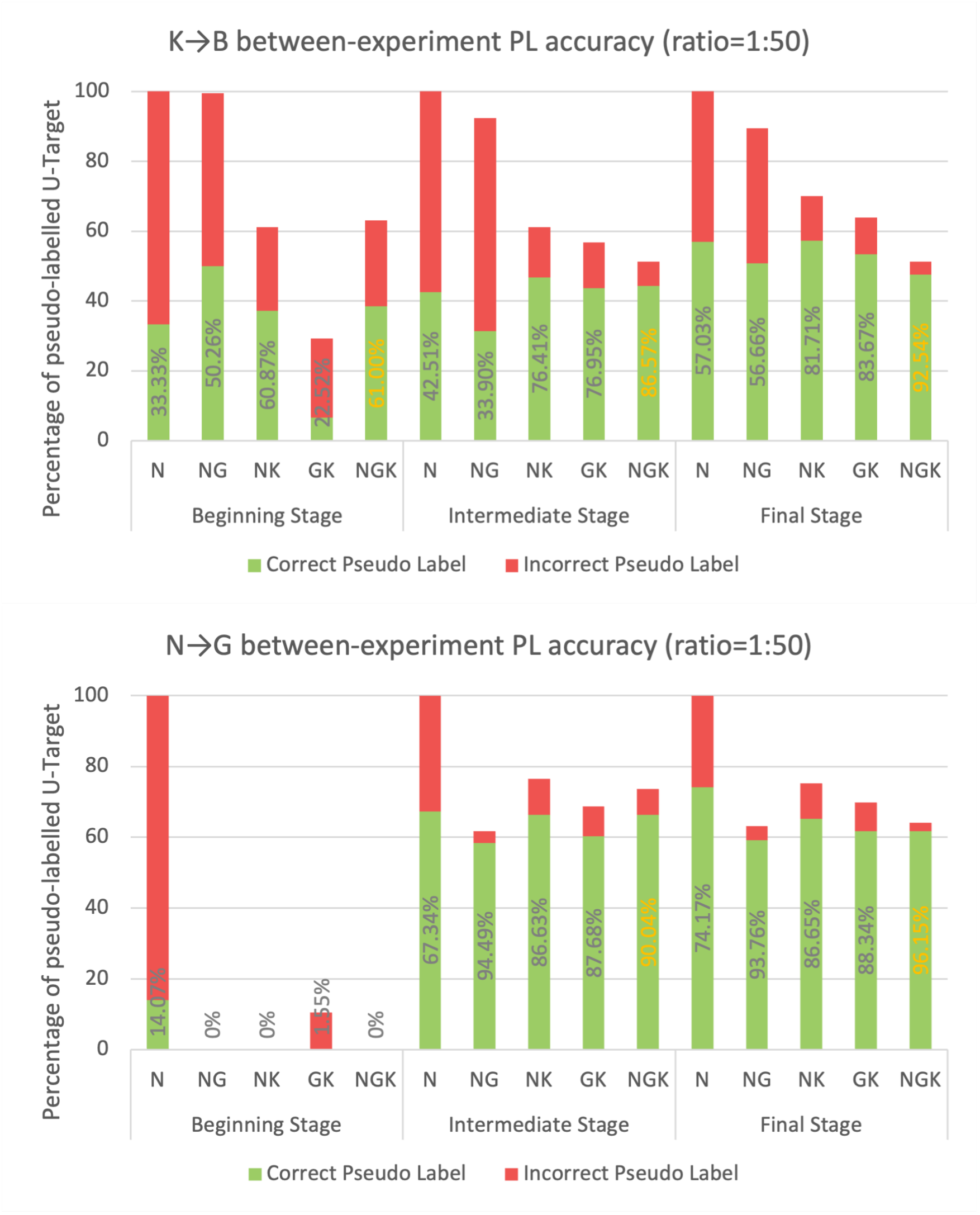}
    \caption{Pseudo-label accuracy under default data scarcity ratio between ablated experiments}
    \label{fig:label_accuracy_50_b}
    \vspace{-0.5cm}
\end{figure}

\subsection{Performance Evaluation}\label{sec:performance_evaluation}

\textbf{Performance Analysis under Default Data Scarcity Ratio}: We analyse the performance against state-of-the-art counterparts under the default target data scarcity ratio. As indicated in Table. \ref{tab:grand_result_table}, the GGA clearly outperforms all comparing methods by at least $4.2\%$. The best-performed comparing method WCGN utilises graph learning framework, however, it does not perform well under the data-scarce IoT scenario, its PL assignment strategy also \rw{lacks} consideration of geometric properties and \rw{assigns} PL in an isolated manner. Hence, it is natural to observe a performance boost achieved by the GGA. 

\textbf{Performance Analysis under Diverse Data Scarcity Ratios}: To verify the effectiveness and robustness of the GGA under varied target data scarcities, we vary the data scarcity ratio $n_{TL}:n_{TU}$ between $1:10$ and $1:100$. Following \cite{li2020simultaneous,yao2019heterogeneous,ning2021malware}, the $1:100$ case is enough to represent an extreme data-scarce setting. We randomly pick three tasks and \rw{present their performance in Table. \ref{tab:ratio_performance_table}. }As we can observe, the GGA achieves the best intrusion detection performance \rw{among all three tasks} under all data scarcity settings. It yields a $4.36\%$ and $8.29\%$ overall average performance increase compared with the best and second best-performed method WCGN and STN. Moreover, under the extreme data scarcity case, the performance boost achieved by GGA reaches $4.6\%$ and $8.63\%$ compared with the best and second best-performed counterpart WCGN and DDAC, and only presents a $0.87\%$ drop compared with the performance under the $1:50$ case, which further verifies the superiority and robustness of the GGA when working under diverse data scarcity conditions. 

\textbf{Performance Analysis using Diverse Evaluation Metrics}: To further verify the effectiveness of the GGA method using evaluation metrics other than accuracy, we randomly select $2$ tasks and record the performance using another $4$ evaluation metrics, and present the result in Table. \ref{tab:metrics_performance_table}. We observe GGA achieves superior performance on all evaluation metrics. Specifically, the highest precision indicates that most of the flagged malicious decisions made by GGA are correct. The highest recall means the GGA can flag the most amount of intrusions among all malicious traffic. As a harmonic mean of precision and recall, the highest F1-score indicates that the GGA can balance properly between flagging malicious actions and avoiding triggering false alarms. Finally, the highest AUC shows the GGA can effectively distinguish malicious intrusions from normal traffic. Together, these evaluation metrics verify the effectiveness of the GGA method and its real-world usability in terms of false alarm avoidance. 

\rw{\textbf{Intrusion Detection Performance Summary}}: We verify the GGA has the best performance on all tasks when evaluated using all metrics. Therefore, it is sufficient \rw{to indicate} that the GGA method can accurately flag malicious traffic while not causing severe false alarms. The best F1-score and AUC score performance also verify the GGA has the best ability to distinguish \rw{benign traffic} and different intrusions. Having such capability promotes the real-world usefulness of GGA when performing effective intrusion detection.

\begin{figure}[t]
    \centering
    \includegraphics[width=0.97\columnwidth]{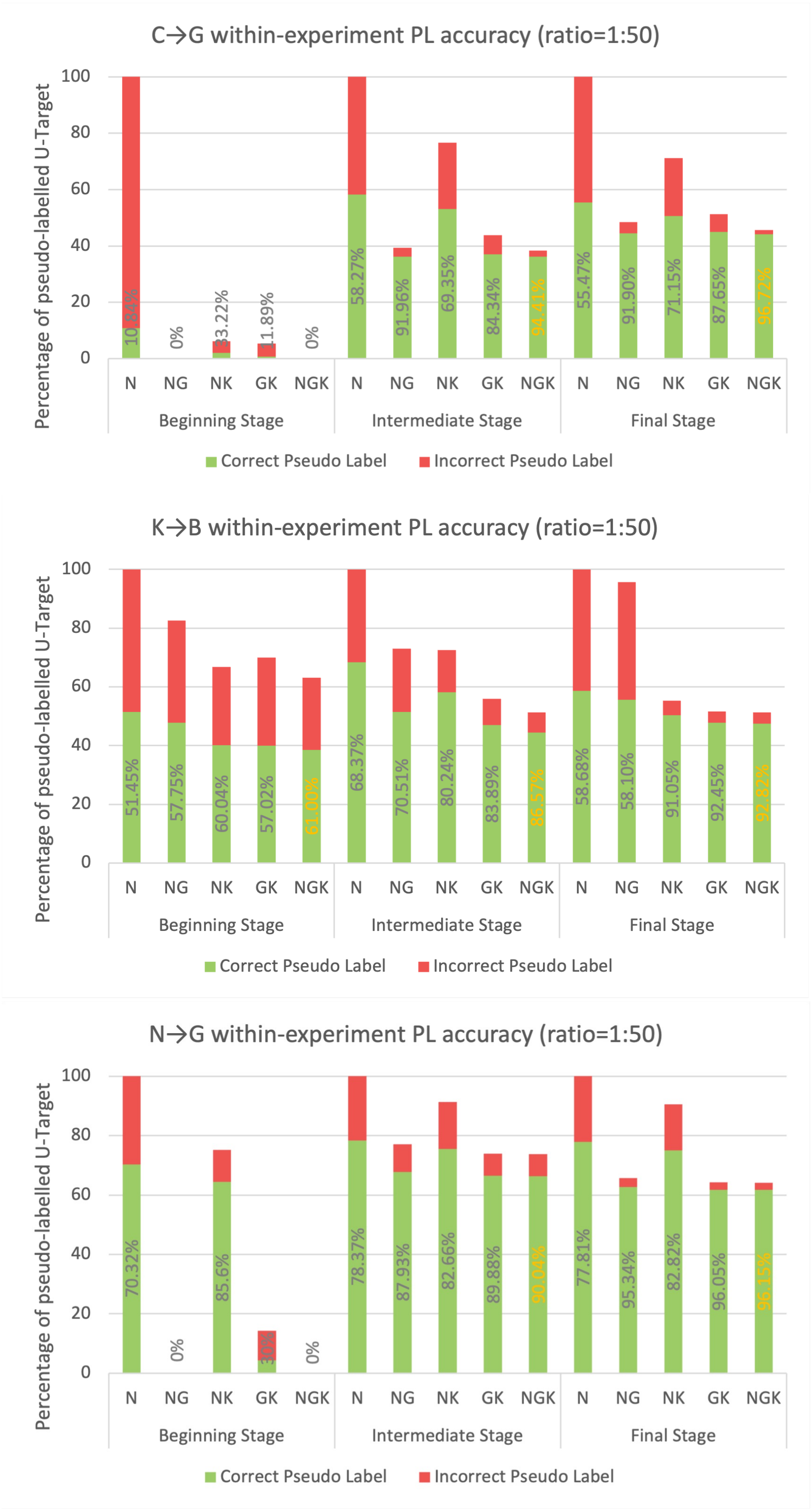}
    \caption{Pseudo-label accuracy under default data scarcity ratio within a single experiment that utilises full PLE}
    \label{fig:label_accuracy_50_w}
    \vspace{-0.5cm}
\end{figure}

\begin{figure}[t]
    \centering
    \includegraphics[width=0.97\columnwidth]{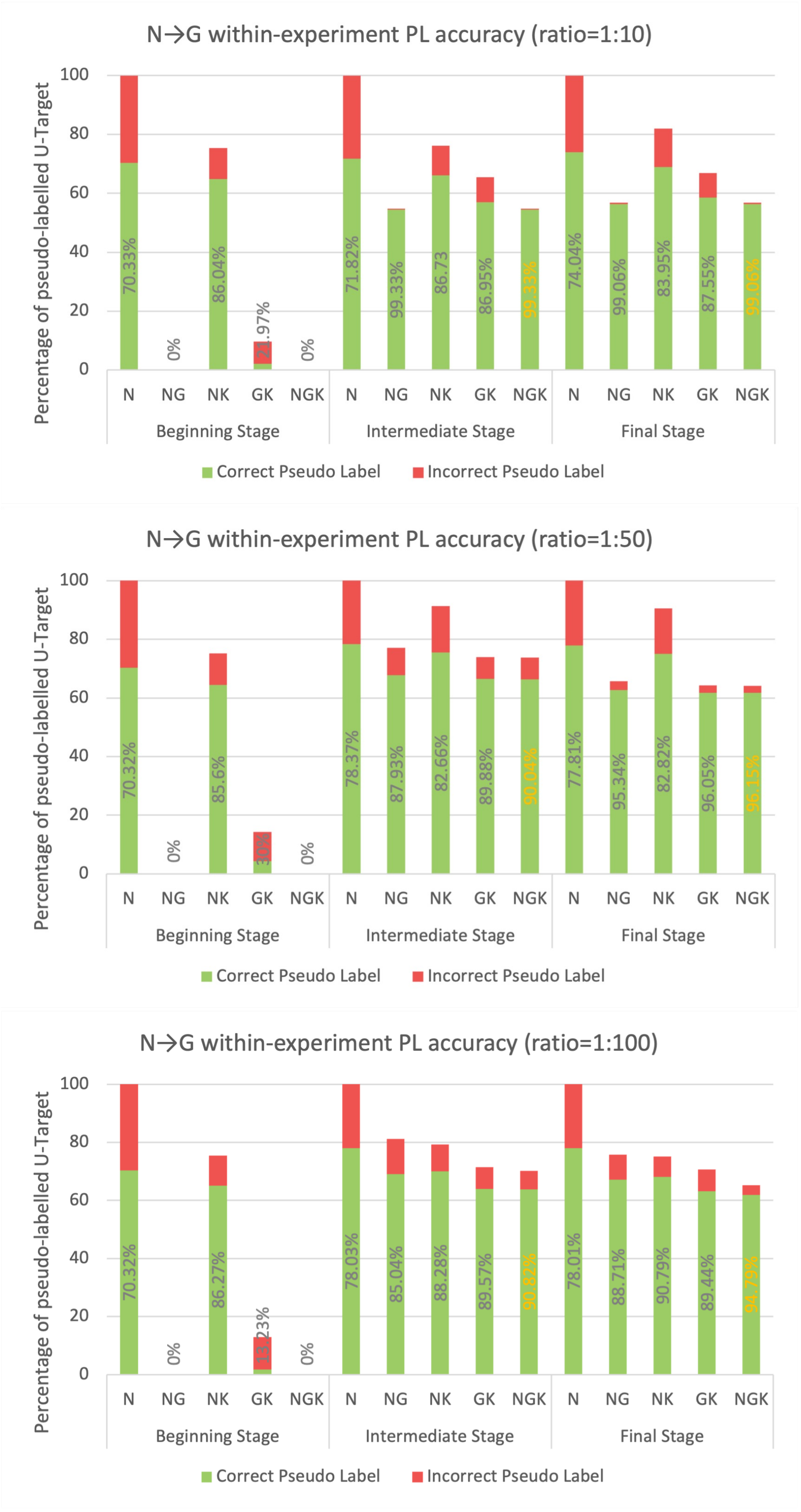}
    \caption{Within-experiment Pseudo-label accuracy under varied data scarcity ratios}
    \label{fig:label_accuracy_varied_ratio_w}
    \vspace{-0.5cm}
\end{figure}

\subsection{Pseudo-label Accuracy Analysis}\label{sec:pseudo_label_accuracy_analysis}

To justify the efficacy of the pseudo-label election (PLE) mechanism, we perform PL accuracy analysis in three ways: (a) analyse the PL accuracy between ablated experiments under default data scarcity ratio; (b) analyse the PL accuracy of different PLE configurations in a single full PLE setting under default data scarcity ratio; (c) perform (b) under varied data scarcity ratios to verify the robustness of the PLE. 

The results on $2$ randomly selected tasks for case (a) has been illustrated in Fig. \ref{fig:label_accuracy_50_b}. Note that N, G and K represents NN prediction vote, geometric property vote and neighbourhood vote, respectively. The height of each bar represents percentage of unlabelled target data being pseudo-labelled, and red-green colour and the value written on each bar indicates the accuracy of PL assignment. As we can observe, using the full PLE yields advantages in three-fold: (1) the full PLE achieves the highest PL accuracy during all training stages. During the beginning stage, to avoid blindly generating a vast amount of false PLs and mislead the alignment process, the full PLE can even temporarily generate no pseudo-labels, since generating PL blindly can only deteriorate the alignment process. (2) the full PLE can reach a relatively high PL accuracy around $86.6\% - 90\%$ even at the intermediate training stage, which guides the aligning process positively by fully exploiting the unlabelled target data. (3) the full PLE eventually achieves a PL accuracy around $92.5\% - 96.2\%$, which indicates the superiority of PLE on accurate PL assignment. Although the full PLE may not generate the highest amount of PLs, however, the accuracy matters more than the amount, as indicated by the superior performance achieved by the full PLE during the ablation study. 

Besides the pseudo-label accuracy analysis performed between ablated experiments, we also perform the PL accuracy analysis within a single full PLE experiment under $3$ randomly picked tasks. As indicated in Fig. \ref{fig:label_accuracy_50_w}, the within-experiment \rw{performances also comply} with the advantages summarised above. The full PLE can stably achieve the highest PL assignment accuracy during all training stages. During each stage, ablating any PLE constituting component will cause the PL accuracy to drop significantly. This result further verifies the usability of all components considered in the PLE. 

To demonstrate the robustness of the PLE under varied data scarcities, the within-experiment PL accuracy is analysed under varied data scarcity ratios as indicated in Fig. \ref{fig:label_accuracy_varied_ratio_w}. Under a relatively low data scarcity case, the PLE can achieve a $99.06\%$ PL accuracy during the final training \rw{stage}. Even under the extreme data scarcity case, the PL accuracy only drops by $4.27\%$ compared with the $1:10$ case, which demonstrates that under varied data scarcities, the PLE can work robustly to generate accurate PL assignment and benefit positive intrusion knowledge transfer during the geometric graph alignment process.

\subsection{Ablation Study}\label{sec:ablation_study}

{\renewcommand{\arraystretch}{1.2}%
\begin{table*}[!ht]
    \centering
    \caption{Ablation Study Group A: Intrusion detection accuracy of GGA with ablated GGA components}
    \label{tab:ablation_group_a}
    \begin{tabular}{c|cccc|ccc|c}
    \Xhline{2\arrayrulewidth}
    \multirow{2}{*}{Experiment} & \multicolumn{4}{c|}{GGA Components} & \multicolumn{3}{c|}{Task} & \multirow{2}{*}{Avg} \\ \cline{2-8}
     & Shape Keeping & Rotation Avoidance & Centre Pt Match & Vertex Semantic Match & K $\rightarrow$ B & N $\rightarrow$ G & C $\rightarrow$ W &  \\ \hline
    A$_1$ & \xmark\hspace{1mm}($\gamma_{min},\gamma_{max} = 0$) & \checkmark & \checkmark & \checkmark & 71.83 & 74.2 & 56.64 & 67.56 \\
    A$_2$ & \checkmark & \xmark\hspace{1mm}($\eta = 0$) & \checkmark & \checkmark & 66.75 & 77.42 & 56.36 & 66.84 \\
    A$_3$ & \checkmark & \checkmark & \xmark\hspace{1mm}($\lambda = 0$) & \checkmark & 69.07 & 75.47 & 57.74 & 67.43 \\
    A$_4$ & \checkmark & \checkmark & \checkmark & \xmark\hspace{1mm}($\alpha = 0$) & 75.63 & 77.77 & 57.86 & 70.42 \\ \hline
    \rowcolor{gray}
    \textbf{Full GGA} & \checkmark & \checkmark & \checkmark & \checkmark & \textbf{77.26} & \textbf{79.18} & \textbf{58.71} & \textbf{71.72} \\ \Xhline{2\arrayrulewidth}
    \end{tabular}
\end{table*}}

{\renewcommand{\arraystretch}{1.2}%
\begin{table*}[!ht]
    \centering
    \caption{Ablation Study Group B: Intrusion detection accuracy of GGA with ablated PLE components}
    \label{tab:ablation_group_b}
    \begin{tabular}{c|ccc|ccc|c}
    \Xhline{2\arrayrulewidth}
    \multirow{2}{*}{Experiment} & \multicolumn{3}{c|}{Pseudo Label Election Mechanism Components} & \multicolumn{3}{c|}{Task} & \multirow{2}{*}{Avg} \\ \cline{2-7}
     & NN Label & Geometric Label & Neighbourhood Label & K $\rightarrow$ B & N $\rightarrow$ G & C $\rightarrow$ W &  \\ \hline
    B$_1$ & \checkmark & \xmark & \xmark & 72.62 & 74.37 & 56.01 & 67.67 \\
    B$_2$ & \checkmark & \checkmark & \xmark & 75.26 & 75.17 & 56.72 & 69.05 \\
    B$_3$ & \checkmark & \xmark & \checkmark & 73.76 & 77.25 & 57.78 & 69.60 \\
    B$_4$ & \xmark & \checkmark & \checkmark & 69.48 & 76.75 & 56.42 & 67.55 \\ \hline
    \rowcolor{gray}
    \textbf{Full GGA} & \checkmark & \checkmark & \checkmark & \textbf{77.26} & \textbf{79.18} & \textbf{58.71} & \textbf{71.72} \\ \Xhline{2\arrayrulewidth}
    \end{tabular}
\end{table*}}

{\renewcommand{\arraystretch}{1.2}%
\begin{table}[!t]
    \centering
    \caption{Ablation Study Group C: Intrusion detection accuracy of methods with different graph alignment mechanisms}
    \label{tab:ablation_group_c}
    \begin{tabular}{c|p{1.6cm}|p{0.8cm}p{0.8cm}p{0.8cm}|p{0.8cm}}
    \Xhline{2\arrayrulewidth}
    \multirow{2}{*}{Experiment} & Alignment & \multicolumn{3}{c|}{Task} & \multirow{2}{*}{Avg} \\ \cline{3-5}
     & Mechanism & K$\rightarrow$B & N$\rightarrow$G & C$\rightarrow$W &  \\ \hline
    C & Vertex E-dist Alignment & 74.38 & 74.79 & 57.65 & 68.94 \\ \hline
    \rowcolor{gray}
    \textbf{Full GGA} & \textbf{Geometric Graph Alignment} & \textbf{77.26} & \textbf{79.18} & \textbf{58.71} & \textbf{71.72} \\ \Xhline{2\arrayrulewidth}
    \end{tabular}
\end{table}}

We further investigate the efficacy of GGA's constituting components. Ablation group A \rw{has the corresponding GGA components} in Equation \ref{equ:overall_objective} being turned off. Ablation group B \rw{has} different PLE voters being ablated. Ablation group C compares GGA with the method that uses direct vertex Euclidean distance alignment as an alternative, which is defined as follows: 
\begin{equation}
    \begin{split}
        & \min \sum_{i}^{K} \sum_{(A, B)} ||V_{A}^i - V_{B}^i||^2, \\
        & (A, B) \in \{(S, TL+PL), (S, S+TL+PL), \\
        & (TL+PL, S+TL+PL)\}\,,
    \end{split}
\end{equation}
where $S+TL+PL$ stands for combining instances from the source, labelled-target and pseudo-labelled target domains. 

As indicated in Table. \ref{tab:ablation_group_a} to Table. \ref{tab:ablation_group_c}, the full GGA outperforms all its ablated counterparts by $3.4\%$ on average, \rw{verifying} positive contributions made by all constituting components towards a fine-grained geometric graph alignment. In ablation group A, the rotation avoidance mechanism is the best performance contributor with $4.9\%$ of performance boost, followed by the centre matching, shape keeping mechanism and vertex-level semantic preservation. In ablation group B, the results verify that all three voting components are indispensable. The performance will drop by $3.3\%$ on average without any one of them. Finally, in ablation group C, a $2.8\%$ performance reduction is observed by the Euclidean distance-based pure vertex alignment procedure. It is natural to observe since there are huge heterogeneities between domains, therefore, pure vertex-level distance-based alignment may not be strong enough to enforce a fine-grained graph alignment, which results in degraded intrusion knowledge transfer. By jointly considering several granularities from shape keeping to vertex-level alignment, the GGA can facilitate a finer alignment and an enhanced intrusion detection performance.

\begin{figure*}[t]
    \centering
    \includegraphics[width=0.8\textwidth]{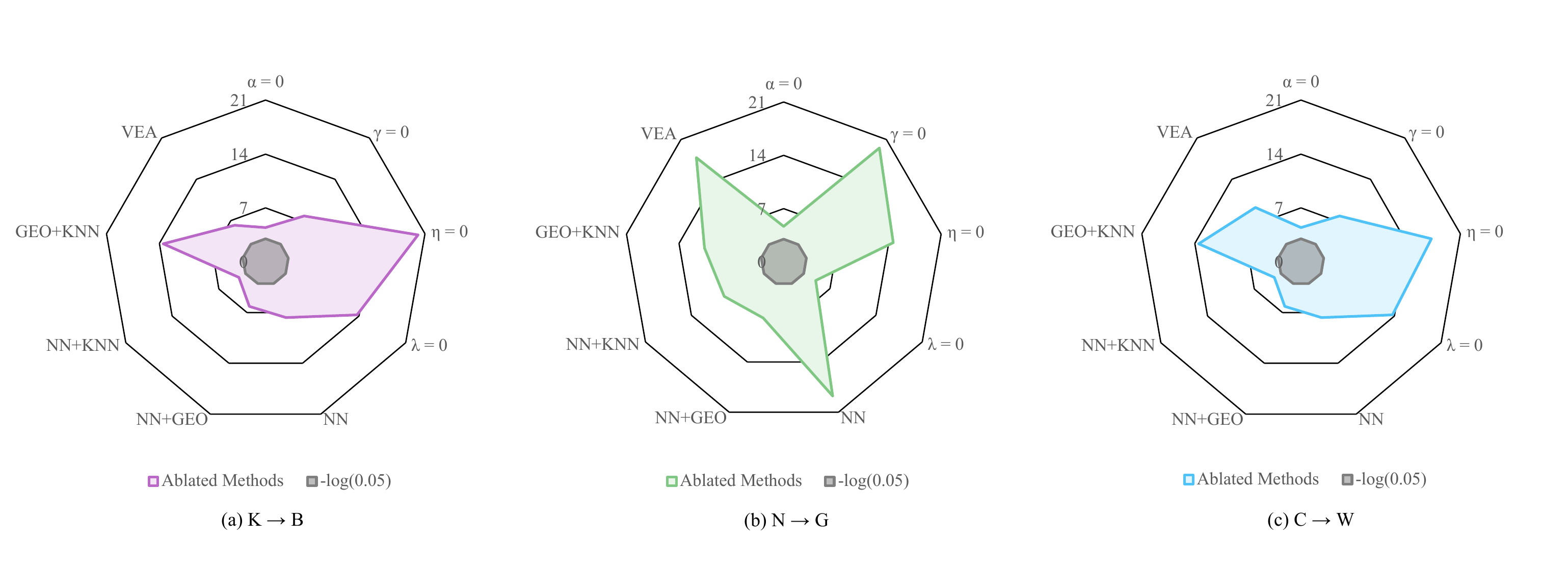}
    \caption{Significance T-tests on $3$ randomly selected tasks have been performed to verify the statistical soundness of the contributions from different constituting components of the GGA. The grey shaded area denotes the significant threshold $-\log(0.05)$. Among each dimension, the wider the coverage is, the more significant the contribution is on that ablated component. }
    \label{fig:ablation_hypothesis_figure}
\end{figure*}

\subsection{Hypothesis Testing for Ablation Study}\label{sec:hypothesis_testing_for_ablation_study}

To verify the statistical significance of the contributions made by each constituting component, i.e., the performance boost is not observed randomly by chance, we perform significance T-test on $3$ randomly selected tasks. As illustrated in Fig. \ref{fig:ablation_hypothesis_figure}, the grey shaded area denotes the significant threshold $-\log(0.05)$. Among each dimension, the wider the coverage is, the more significant the contribution is on that ablated component. As we can see from Fig. \ref{fig:ablation_hypothesis_figure}, the coloured area has wider coverage than the grey shaded area among all dimensions, which indicates that the contributions made by all constituting components have statistical soundness. Therefore, all proposed components are indispensable for GGA to achieve a fine-grained intrusion knowledge transfer via graph alignment.

\begin{figure*}[t]
    \centering
    \includegraphics[width=0.97\textwidth]{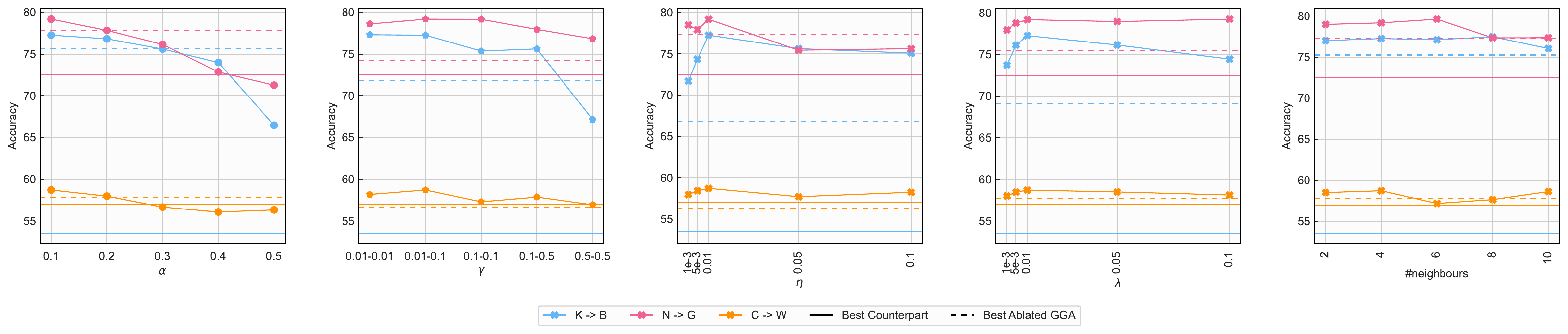}
    \caption{The parameter sensitivity analysis of the GGA method for hyperparameters $\alpha$, $\gamma$, $\eta$, $\lambda$, and $\#neighbour$ under their corresponding reasonable range. As shown in the legend, the solid and dashed horizontal lines indicate the best comparing method and the best-performed GGA ablated counterpart under each task, respectively. }
    \label{fig:paramter_sensitivity_figure}
    \vspace{-0.5cm}
\end{figure*}

\subsection{Parameter Sensitivity}\label{sec:parameter_sensitivity}

The parameter sensitivity of the GGA method has been illustrated in Fig. \ref{fig:paramter_sensitivity_figure}. The GGA shows relatively stable performance under these hyperparameter ranges without showing severe fluctuations. Besides, the GGA outperforms the best-performed comparing method under nearly all hyperparameter ranges. Additionally, the GGA constantly shows superior performance than its best ablated counterpart. Therefore, we verify that the GGA method is robust on varied hyperparameter settings. 

Besides, during all experiments, only a fixed set of hyperparameters is used to tackle diverse data domains. The GGA can constantly show satisfying performance without the need to perform time-consuming hyperparameter resetting. Therefore, it further demonstrates the robustness of GGA on hyperparameters and its usefulness when tackling diverse intrusion data \rw{domains}.

\subsection{Computational Efficiency}\label{sec:computational_efficiency}

{\renewcommand{\arraystretch}{1.2}%
\begin{table}[!ht]
    \vspace{-3mm}
    \caption{Comparison on training time per epoch (second)}
    \centering
    \label{tab:training_time_per_epoch}
    \begin{tabular}{c|ccc|c}
    \Xhline{2\arrayrulewidth}
    S $\rightarrow$ T & K $\rightarrow$ B & N $\rightarrow$ G & C $\rightarrow$ W & Avg \\ \hline
    DDAC & 7.54 & 3.86 & 1.96 & 4.45 \\
    WCGN & 0.18 & \textbf{0.11} & 0.15 & 0.15 \\ \hline
    \rowcolor{gray}
    \textbf{GGA (Ours)} & \textbf{0.17} & \textbf{0.11} & \textbf{0.13} & \textbf{0.14} \\ \Xhline{2\arrayrulewidth}
    \end{tabular}
\end{table}}

{\renewcommand{\arraystretch}{1.2}%
\begin{table}[!ht]
    \caption{Comparison on inference time per instance (microsecond $=10^{-6}$ second)}
    \centering
    \label{tab:inference_time_per_instance}
    \begin{tabular}{c|ccc|c}
    \Xhline{2\arrayrulewidth}
    S $\rightarrow$ T & K $\rightarrow$ B & N $\rightarrow$ G & C $\rightarrow$ W & Avg \\ \hline
    DDAC & 1.43 & 1.76 & 0.96 & 1.38 \\
    WCGN & 0.20 & 0.17 & 0.19 & 0.19 \\ \hline
    \rowcolor{gray}
    \textbf{GGA (Ours)} & \textbf{0.17} & \textbf{0.14} & \textbf{0.17} & \textbf{0.16} \\ \Xhline{2\arrayrulewidth}
    \end{tabular}
\end{table}}

Finally, to verify the computational efficiency of the GGA method, we measure both the training time per epoch and inference time per instance, and make comparison between two best-performed baseline counterparts. The results are presented in Table \ref{tab:training_time_per_epoch} and \ref{tab:inference_time_per_instance}. As we can observe, the GGA achieves the best training and inference efficiency. Specifically, the GGA trains $31$ times and $6.67\%$ faster than DDAC and WCGN, respectively. Besides, the GGA also achieves the fastest inference speed, which outperforms the best-performed counterpart WCGN by $15.79\%$. Hence, it verifies the efficiency of the GGA, making it suitable to be used under computationally-constrained IoT scenarios.

\section{Conclusion}\label{sec:conclusion}

In this paper, we utilise the knowledge rich network intrusion domain to facilitate accurate intrusion detection for data-scarce IoT domain. We tackle this HDA problem through a geometric graph alignment approach. Firstly, a pseudo-label election mechanism is employed to exploit the unlabelled target instances, which jointly considers the network prediction, geometric property and neighbourhood information to boost the PL assignment accuracy. The PLE avoids geometrically diverged confident but wrong PLs and near-boundary ambiguous PLs. Then, both intrusion domains are formulated as graphs, with the GGA performed using four mechanisms, from general graph granularity to vertex-level alignment. Specifically, the graph shape is kept via a confused discriminator that is incapable to distinguish the origin of weighted adjacency matrices. Besides, the rotation avoidance mechanism and the centre point matching mechanism avoid graph misalignment caused by rotation and symmetry, respectively. Additionally, the vertex-level semantic is preserved to facilitate a more fine-grained graph alignment. By forming these mechanisms into a holistic whole, the GGA method can align intrusion graphs in a fine-grained manner, which benefits the intrusion knowledge transfer between domains. Comprehensive experiments demonstrate the state-of-the-art performance of the GGA method. Insight analyses also verify the usefulness of each constituting component of the GGA method.

%%%%%%%%%%%%%%%%%%%%%%
%%%%%%%%%%%%%%%%%%%%%%
%% Acknowledgements %%
%%%%%%%%%%%%%%%%%%%%%%
%%%%%%%%%%%%%%%%%%%%%%

\ifCLASSOPTIONcompsoc
  \section*{Acknowledgments}
\else
  \section*{Acknowledgment}
\fi

This work is supported by the Third Xinjiang Scientific Expedition Program (Grant No.2021xjkk1300), and also in part by Science and Technology Development Fund of Macao SAR (FDCT) (1058No.0015/2019/AKP), Chinese Academy of Sciences President’s International Fellowship Initiative (Grant No. 2023VTA0001). 

\ifCLASSOPTIONcaptionsoff
  \newpage
\fi

%%%%%%%%%%%%%%%%%%%%%%%%%%%%%
%%%%%%%%%%%%%%%%%%%%%%%%%%%%%
%% Bibliography References %%
%%%%%%%%%%%%%%%%%%%%%%%%%%%%%
%%%%%%%%%%%%%%%%%%%%%%%%%%%%%

\bibliographystyle{IEEEtran}
\bibliography{GGA}

%%%%%%%%%%%%%%%
%%%%%%%%%%%%%%%
%% Biography %%
%%%%%%%%%%%%%%%
%%%%%%%%%%%%%%%

\vspace{-1cm}
\begin{IEEEbiography}[{\includegraphics[width=1in,height=1.25in,clip,keepaspectratio]{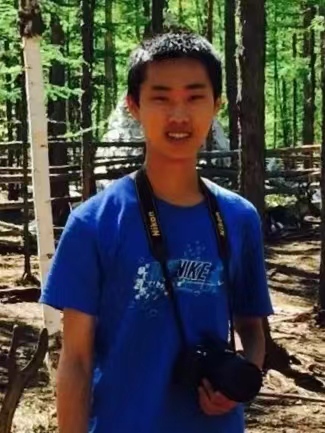}}]{Jiashu Wu} received BSc. degree in Computer Science and Financial Mathematics \& Statistics from the University of Sydney, Australia (2018), and M.IT degree in Artificial Intelligence from the University of Melbourne, Australia (2020). He is currently pursuing his Ph.D. at the University of Chinese Academy of Sciences (Shenzhen Institute of Advanced Technology, Chinese Academy of Sciences). His research interests including big data and cloud computing. 
\end{IEEEbiography}
\vspace{-1.9cm}
\begin{IEEEbiography}[{\includegraphics[width=1in,height=1.20in,clip,keepaspectratio]{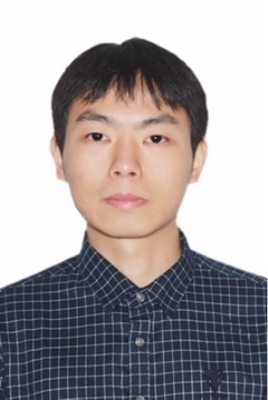}}]{Hao Dai} received the BS and M.Sc degrees in Communication and Electronic Technology from the Wuhan University of Technology in 2015 and 2017, respectively. He is currently working toward the Ph.D. degree in the Shenzhen Institute of Advanced Technology, Chinese Academy of Sciences. His research interests include mobile edge computing, federated learning and deep reinforcement learning. 
\end{IEEEbiography}
\vspace{-1.9cm}
\begin{IEEEbiography}[{\includegraphics[width=1in,height=1.25in,clip,keepaspectratio]{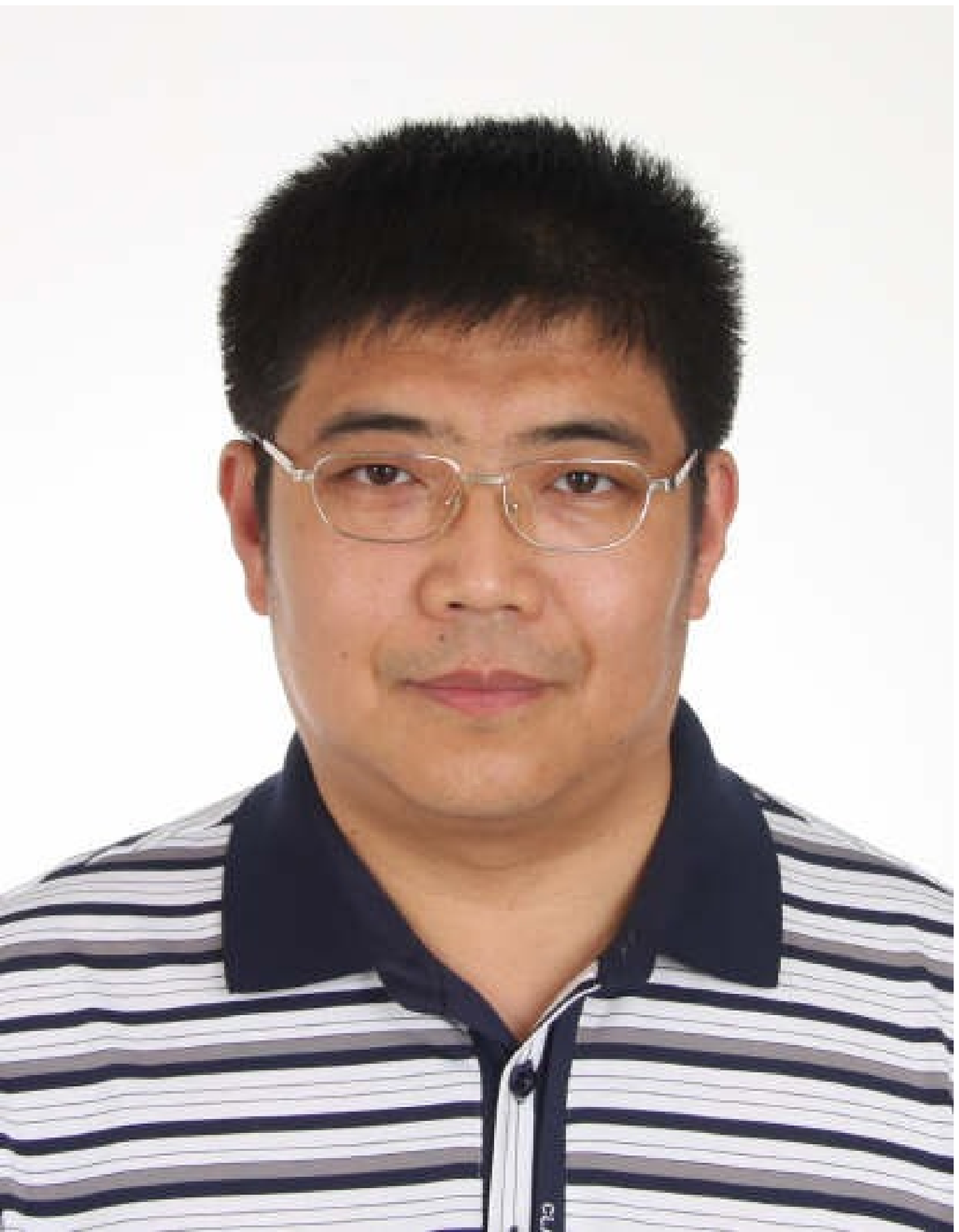}}]{Yang Wang} received the BSc degree in applied mathematics from Ocean University of China (1989), and the MSc. and PhD. degrees in computer science from Carlton University (2001) and University of Alberta, Canada (2008), respectively. He is currently with Shenzhen Institutes of Advanced Technology, Chinese Academy of Sciences, as a professor and with Xiamen University as an adjunct professor. His research interests include service and cloud computing, programming language implementation, and software engineering. He is an Alberta Industry R\&D Associate (2009-2011), and a Canadian Fulbright Scholar (2014-2015). 
\end{IEEEbiography}
\vspace{-1.9cm}
\begin{IEEEbiography}[{\includegraphics[width=1in,height=1.25in,clip,keepaspectratio]{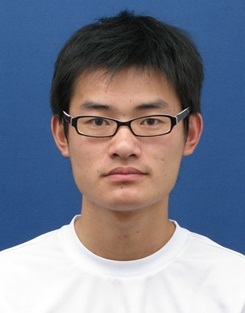}}]{Kejiang Ye}
received his BSc. and Ph.D. degree in Computer Science from Zhejiang University in 2008 and 2013 respectively. He was also a joint PhD student at University of Sydney from 2012 to 2013. After graduation, he worked as Post-Doc Researcher at Carnegie Mellon University from 2014 to 2015 and Wayne State University from 2015 to 2016. He is currently an Associate Professor at Shenzhen Institutes of Advanced Technology, Chinese Academy of Science. His research interests focus on the performance, energy, and reliability of cloud computing and network systems.
\end{IEEEbiography}
\vspace{-1.9cm}
\begin{IEEEbiography}[{\includegraphics[width=1in,height=1.20in,clip,keepaspectratio]{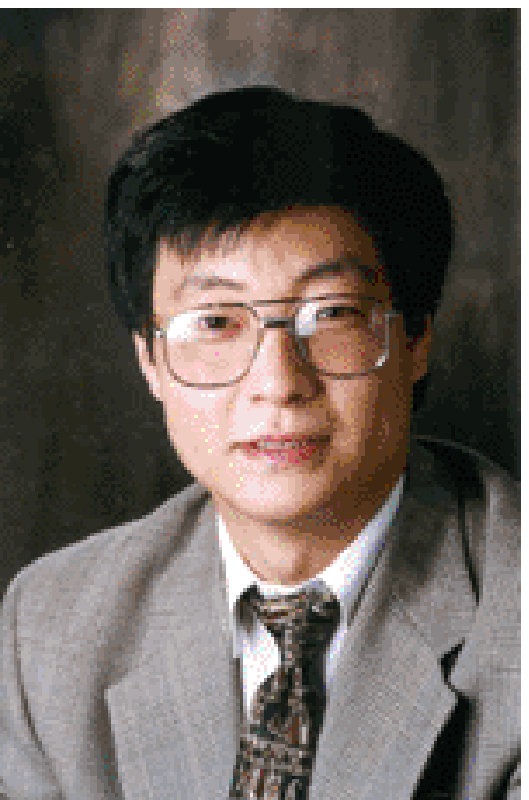}}]{Chengzhong Xu} received the Ph.D. degree from the University of Hong Kong in 1993. He is currently the Dean of Faculty of Science and Technology, University of Macau, China, and the Director of the Institute of Advanced Computing and Data Engineering, Shenzhen Institutes of Advanced Technology of Chinese Academy of Sciences.His research interest includes parallel and distributed systems, service and cloud computing, and software engineering. He has published more than 200 papers in journals and conferences. He serves on a number of journal editorial boards, including IEEE TC, IEEE TPDS, IEEE TCC, JPDC and China Science Information Sciences. He is a fellow of the IEEE.
\end{IEEEbiography}
\vspace{-1.5cm}

\clearpage

%%%%%%%%%%%%%%%%%%%%%%%%%%%%%%%%%%%%%%%%%%%%%%%%%%%%%%%%%%%%%%%%%%%%%%%%%%%%%%%
%%%%%%%%%%%%%%%%%%%%%%%%%%%%%%%%%%%%%%%%%%%%%%%%%%%%%%%%%%%%%%%%%%%%%%%%%%%%%%%
%%%%%%%%%%%%%%%%%%%%%%%%%%%%%%%%%%%%%%%%%%%%%%%%%%%%%%%%%%%%%%%%%%%%%%%%%%%%%%%
%%% Appendix Appendix Appendix Appendix Appendix Appendix Appendix Appendix %%%
%%% Appendix Appendix Appendix Appendix Appendix Appendix Appendix Appendix %%%
%%% Appendix Appendix Appendix Appendix Appendix Appendix Appendix Appendix %%%
%%%%%%%%%%%%%%%%%%%%%%%%%%%%%%%%%%%%%%%%%%%%%%%%%%%%%%%%%%%%%%%%%%%%%%%%%%%%%%%
%%%%%%%%%%%%%%%%%%%%%%%%%%%%%%%%%%%%%%%%%%%%%%%%%%%%%%%%%%%%%%%%%%%%%%%%%%%%%%%
%%%%%%%%%%%%%%%%%%%%%%%%%%%%%%%%%%%%%%%%%%%%%%%%%%%%%%%%%%%%%%%%%%%%%%%%%%%%%%%

\section{Appendix}

\textbf{Acronym Table}: For better readability, we provide the following acronym table for reference. Note that all acronyms are defined in the main text \rw{as well as} at the first time they are introduced. 

{\renewcommand{\arraystretch}{1.2}%
\begin{table}[!ht]
    \vspace{-5mm}
    \caption{Acronym table}
    \centering
    \label{tab:acronym_table}
    \begin{tabular}{l|l}
    \Xhline{2\arrayrulewidth}
    Acronym & Interpretation \\ \hline
    IID & IoT Intrusion Detection \\
    NID & Network Intrusion Detection \\
    GGA & Geometric Graph Alignment \\
    DA & Domain Adaptation \\
    NI & Network Intrusion \\
    II & IoT Intrusion \\
    PL(s) & Pseudo Label(s) \\
    HDA & Heterogeneous Domain Adaptation \\
    PLE & Pseudo Label Election \\
    WAM(s) & Weighted Adjacency Matrix(Matrices) \\
    NN & Neural Network \\ \Xhline{2\arrayrulewidth}
    \end{tabular}
\end{table}}

\textbf{Notation Table}: We provide a notation table for better readability. 

{\renewcommand{\arraystretch}{1.2}%
\begin{table}[!ht]
    \caption{Notation table}
    \centering
    \label{tab:notation_table}
    \rw{\begin{tabular}{l|p{7cm}}
    \Xhline{2\arrayrulewidth}
    Notation & Interpretation \\ \hline
    $\mathcal{D}_{*}$ & Domain, $* \in \{S, TL, TU\}$ \\
    $x_{*_i}$ & The $i$\textsuperscript{th} instance of domain $*$ \\
    $y_{*_i}$ & The intrusion class of the $i$\textsuperscript{th} instance of domain $*$ \\
    $n_{*}$ & Number of instances in domain $*$ \\
    $d_{*}$ & Dimension of domain $*$ instances \\
    $K$ & Number of intrusion categories \\
    $G_{X}$ & Domain graph of domain $X$ \\
    $V_{X}$ & Vertices in domain graph $G_{X}$ \\
    $E_{X}$ & Edges in domain graph $G_{X}$ \\
    $V_{X}^i$ & Class $i$ vertices in domain graph $G_{X}$ \\
    $E_{X}$ & Feature projector for domain X \\
    $f(x_i)$ & Features projected by the feature projector \\ 
    $PL_{G}^i$ & The geometric label for the $i$\textsuperscript{th} unlabelled target instance \\
    $\mathcal{X}_{S}^{(k)}$ & Source domain instances belong to class $k$ \\
    $\mu_{S+TL}^{(k)}$ & Centroid of the $k$\textsuperscript{th} class of source and labelled target domain instances \\
    $M_X$ & The weighted adjacency matrix of domain $X$ \\
    $\mathcal{L}_{SK}$ & Shape keeping loss \\
    $D()$ & The discriminator \\
    $\mathcal{L}_{R}$ & Rotation avoidance loss \\
    $\mathcal{L}_{CP}$ & Centre point matching loss \\
    $q^{(k)}$ & The correlation semantic of class $k$ for the source domain \\
    $C()$ & The shared classifier \\
    $T$ & Temperature hyperparameter controlling the semantic smoothness \\
    $p_i$ & The correlation semantic of the $i$\textsuperscript{th} labelled target instance \\
    $\mathcal{L}_{V}$ & Vertex-level alignment loss \\
    $\alpha$ & Hyperparameter in $\mathcal{L}_{V}$ \\
    $\mathcal{L}_{ce}$ & Cross entropy loss \\
    $\mathcal{L}_{SUP}$ & The supervision loss of source domain \\
    $\gamma$ & Hyperparameter balancing the weight of $\mathcal{L}_{SK}$ \\
    $\eta$ & Hyperparameter balancing the weight of $\mathcal{L}_{R}$ \\
    $\lambda$ & Hyperparameter balancing the weight of $\mathcal{L}_{CP}$ \\
    $TP^{(k)}$ & True positive for intrusion category $k$ \\ \Xhline{2\arrayrulewidth}
    \end{tabular}}
    \vspace{-3mm}
\end{table}}

\end{document}